\documentclass[showpacs,prb,manuscript]{revtex4}
\usepackage{amsmath}
\usepackage{amssymb}
\usepackage{epsfig}

\begin{document}

\title{Elastic Spin Relaxation Processes in Semiconductor Quantum Dots}
\author{Y. G. Semenov and K. W. Kim}

\address{Department of Electrical and Computer Engineering\\
North Carolina State University, Raleigh, NC 27695-7911}

\begin{abstract}
Electron spin decoherence caused by elastic spin-phonon processes
is investigated comprehensively in a zero-dimensional environment.
Specifically, a theoretical treatment is developed for the
processes associated with the fluctuations in the phonon potential
as well as in the electron procession frequency through the
spin-orbit and hyperfine interactions in the semiconductor quantum
dots. The analysis identifies the conditions (magnetic field,
temperature, etc.) in which the elastic spin-phonon processes can
dominate over the inelastic counterparts with the electron
spin-flip transitions. Particularly, the calculation results
illustrate the potential significance of an elastic decoherence
mechanism originating from the intervalley transitions in
semiconductor quantum dots with multiple equivalent energy minima
(e.g., the $X$ valleys in SiGe). The role of lattice anharmonicity
and phonon decay in spin relaxation is also examined along with
that of the local effective field fluctuations caused by the
stochastic electronic transitions between the orbital states.
Numerical estimations are provided for typical GaAs and Si-based
quantum dots.
\end{abstract}

\maketitle


\section{Introduction}

Electron spin relaxation in semiconductor quantum dots (QDs) has received
much attention as they are a natural candidate for the qubit in quantum
computing (QC) (see Refs.~\onlinecite{Cerletti05} and \onlinecite{Valiev05}
for a brief review). Particularly, a significant effort has been devoted to
investigate the spin transition probabilities associated with the spin-flip
interactions, for example, the so-called longitudinal spin relaxation ($%
T_{1} $) (i.e., energy relaxation) induced via the inelastic process of
phonon absorption/emission by an electron spin localized in a QD.
Although generally an importance source of decoherence, the spin-flip events
can sometimes be efficiently suppressed by an appropriate choice of external
conditions including the magnetic field strength or direction. In this case,
alternative mechanisms (such as the elastic spin-phonon scattering that does
not assume energy exchange between the electron Zeeman energy and thermal
bath) can be primarily responsible for decoherence. So far, however, the
elastic processes have received relatively little attention. This disparity
may be due, in part, to the mathematical simplicity in the treatment of the
former case (e.g., calculations based on the Fermi's golden rule), while the
elastic processes of spin decoherence need more sophistical approaches. At
the same time, recent proliferation of the literature on electron spin
relaxation has exacerbated the confusion on this complicated subject matter.
Frequently, an individual study addresses only a specific aspect of spin
decoherence or dephasing without elucidating its place in the general theory
or the interrelationships with other relevant works. Consequently, one faces
difficulties each time in determining which existing mechanisms, processes,
or theories are relevant for spin relaxation under the given external
conditions and specific QD properties.

The purpose of the present study is to systematically identify and analyze
the crucial transversal ($T_2$) spin relaxation processes in the context of
qubit dephasing in the electron spin based QC. Specifically, the role of
spin-phonon interaction on the elastic decoherence processes is examined
theoretically from the perspective of the fluctuations in the effective
magnetic field. The calculation identifies the conditions in which the
elastic spin-lattice processes can dominate over the inelastic counterparts.
The analysis is applicable to the electron spin in both group IV and III-V
QDs. Note that the related issues from the spintronics point of view were
surveyed in Refs.~\onlinecite{Zutic04} and \onlinecite{Bronold04}.

The rest of this paper is organized as follows. Section~II provides a
systematic classification of most, if not all, of the processes identified
so far that lead to a type of electron spin relaxation (i.e., dephasing,
decoherence or energy relaxation). Then, the discussion focuses on those
caused by the spin-phonon interaction and describes different manifestations
of the spin-phonon interaction in a manner consistent to both elastic and
inelastic processes. Subsequently, the elastic decoherence processes that
warrant further analysis are recognized and their potential significance
outlined in the context of QC. Section~III summarizes a theoretical
framework necessary for the mathematical description of electron spin
relaxation in terms of quantum kinetic equation.

Each of the following three sections describes a particular elastic process
that can be dominant under a certain condition. In Sec.~IV, we analyze the
effects of spin precession fluctuations due to the irregular phonon phase
disturbances originating from phonon relaxation via the lattice
anharmonicity or imperfections. Such a situation applies not only to real
(thermally activated) phonons but also to zero-point phonons that the
classical picture cannot address. Section~V considers spin phase diffusion
due to the modulations in the longitudinal g-factor and hyperfine
interaction through the phonon-assisted transitions between the lowest
electronic states (without spin-flip). Section~VI concerns electron spin in
a semiconductor QD with multiple equivalent minima (e.g., Si$_{1-x}$Ge$_{x}$%
). As the degeneracy is lifted due to structural inversion asymmetry,
phonon-mediated transitions between the "valley-split" states can be allowed
when the local crystal imperfections are present. This process can be very
significant since the energy separation of the valley-split states is
generally comparable to the thermal energy in the QC operation. Then, the
discussion on the comparative importance of different processes in the
parameter space is provided in Sec.~VII, followed by a brief conclusion at
the end.


\section{Spin relaxation processes and mechanisms}

For a systematic analysis, Table~I shows simple classification of the
relevant spin relaxation processes identified so far. The term "dephasing"
(with the characteristic time $T_{2}^{\ast }$) concerns phase relaxation of
a QC system consisting of multiple qubits, while "decoherence" ($T_2$) is
for a single qubit (i.e., single electron spin in a QD). Hence, decoherence
contributes to dephasing ($T_{2}\geq T_{2}^{\ast }$). The "random local
fields" category denotes the processes that lead to phase diffusion of the
multi-qubit system without causing decoherence in the individual qubits. The
decoherence or transversal relaxation has contributions from both the
elastic and inelastic processes. The latter involves spin flip transitions
leading to the energy (or longitudinal) relaxation of the spin states (i.e.,
$T_1$). Of the spin decoherence processes, those marked by the \# sign in
the table are induced by the spin-phonon interaction. The listed $T_2$
processes correspond to the uncontrollable interactions with the environment
that assumes to be in a thermodynamic equilibrium at temperature $T$.
Inaccuracies in the parameters of the control pulses during QC operations as
well as uncontrollable perturbations introduced with these pulses are beyond
the current description. Table~II summarizes various manifestations of the
spin-phonon mechanism in a manner consistent to both elastic and inelastic
processes. A brief description on each category is given below.

\subsection{Dephasing in random fields}

The fastest relaxation in a quantum computer consisting of a large number of
qubits [$N\sim 1000$ (see Ref. \onlinecite{Valiev05})] is the dephasing
process associated with transversal magnetization loss with the
characteristic time $T_{2}^{\ast }$. Clearly, the electrons experience a
dispersion of spin precession in the presence of a random effective magnetic
field. Even when the magnetic field can be considered a constant spatially
and temporally for a given qubit within the time scale of QC operation, its
random variation across the ensemble of qubits can induce dephasing of the
system. Although some of its effect may be potentially mitigated by QC
algorithms,\cite{Valiev05,Facchi05} the absence of this additional
complexity provides an obvious advantage.

One reason for the dispersion in the precession frequency is the $g$-tensor
variation over the $N$ qubits. Consequently, even in a very pure crystal,
the unavoidable fluctuation in the QD sizes results in the $g$-factor
dispersion $\Delta g$, which subsequently induces dephasing with the rate of
${T_{gt}^{\ast }}^{-1}= \Delta g\mu _{B}B$ (where $\mu _{B}$ is the Bohr
magneton and $B$ the external magnetic field). This mechanism of spin phase
relaxation has not yet been investigated in the zero-dimensional (0D)
structures. A related case of 2D hole localization in a fluctuating
potential was studied in Refs.~\onlinecite{Marie99} and \onlinecite{SBK02}.
In addition, the randomness of spin-orbital coupling in a quantum well due
to the fluctuation of dopant concentration was considered in Ref.~%
\onlinecite{Glazov05}.

Similarly, electron interaction with nuclear spins can induce the spin phase
diffusion due to the random distribution of the nuclear spin states and
their spatial locations.~\cite{Merkulov02,SK03,Braun05} In principle,
replacement of the magnetic isotopes by nonmagnetic counterparts ($I=0$) can
remove or suppress this source of relaxation.~\cite{Tyryshkin03} However, it
may not be a practical option as a significant level of purification is
required to be effective; the dephasing rate depends only weakly on the
nuclear spin concentration $n_I$ (i.e., $\propto \sqrt{n_I}$). As indicated
in Table~I, the random fields of magnetic impurities also contribute to the
dephasing process. The last one listed in this category, the magneto-dipole
interaction between the qubits, plays a specific role since it can be
incorporated (thus, eliminated) to the QC algorithm under certain
conditions.~\cite{Georgeot01,Byrd05,Valiev05}

\subsection{Inelastic processes}

Among the inelastic processes in Table~I, the direct spin-flip via one
phonon emission/absorption (see Fig.~1) is the most studied due to its
simplicity in concept as well as the importance at low temperatures.~\cite%
{Khaetskii01,Tahan02,GK03,Golovach04,Fal'ko05,Sherman05} The rate of this
process is proportional to the phonon density of states $g(\omega _{q})[\sim
\omega _{q}^{2}$] at the resonant phonon energy $\hbar \omega _{q}=g\mu
_{B}B $ ($\equiv \hbar \omega $) as well as the square of the spin-phonon
matrix element ($\sim \omega _{q}B^{2})$. As a result, a strong magnetic
field dependence ($\sim B^{5}$) is expected for the direct spin-flip. Thus,
at low magnetic fields, the effect of this process can be significantly
reduced. When $k_{B}T<\hbar \omega $, the influence of the temperature can
be ignored as the phonon emission determines the relaxation rate.

The Orbach process\cite{Orbach61,Finn61} considers the spin-flip events
through successive transitions between the ground and excited orbital states
separated by an energy $\delta _{0}$. For example, an electron first makes a
spin-flip transition to the excited state with opposite spin by absorbing a
phonon with energy $\delta _{0} \mp \hbar \omega $, followed by the return
to the ground state through emission of a phonon with energy $\delta _{0}$.
An alternative channel is to make a transition to the excited state without
spin flip (i.e., $\delta _{0}$) and, then, return to the ground state with
opposite spin (i.e., $\delta _{0} \mp \hbar \omega $). Thus, the net effect
is the electron spin flip with the energy of $\hbar \omega $ (i.e., the
Zeeman splitting) transferred to the thermal bath. Under certain conditions,
the corresponding matrix elements can exceed those for the direct spin-flip
transitions between the Kramers doublet.~\cite{Khaetskii00} It should be
stressed that the Orbach process assumes two one-phonon transitions in
tandem. Hence, the first transition with the energy threshold of $\delta
_{0} \mp \hbar \omega $ or $\delta _{0}$ provides the primary temperature
dependence along with a weak influence of the magnetic field.

The two-phonon process concerns the longitudinal spin relaxation via
two-phonon absorption, emission or inelastic phonon scattering (Raman
process). It can be further divided into three groups; the direct spin$-$%
two-phonon transitions due to the nonlinearity of the spin-phonon
Hamiltonian~\cite{Terrile77} [Fig.~2(a)], the virtual transitions
mediated by one-phonon interactions through the excited
states~\cite{Khaetskii01,GK03} [Fig~2(b)], and the anharmonic Raman
process~\cite{Zevin70,Hernandez71} [Fig.~2(c)]. In the case of
virtual transitions, the integration over phonons includes
the poles at the phonon energy $\hbar \omega _{q}\sim \delta _{0}$ if $%
\delta _{0}<k_{B}\Theta $, where $\Theta $ is the Debye temperature. The
treatment of this singularity (the so-called phonon resonant fluorescence)
beyond the conventional perturbation theory\cite{Heitler57} involves
essentially the finite lifetime $\tau _{R}$ of the excited states that
exactly reproduces the relaxation rate of the Orbach process.~\cite%
{Orbach61,Aminov72} This coincidence/similarity is lifted (i.e. poles can be
ignored as compared with Orbach process) when the finite phonon lifetime $%
\tau _{q}$ is considered since $\tau _{R}^{-1}\ll \tau _{q}^{-1}$. As the
processes involving the inelastic phonon scattering show a strong dependence
on temperature ($T^{7}-T^{11}$),~\cite{Khaetskii01,GK03} it has been mostly
ignored in the consideration of QC that operates at very low temperatures.
However, a recent analysis~\cite{GK03} suggests its potential significance
in the low magnetic field regime due to the weak dependence on the field
strength ($B^{2}$).

The process of directly transferring the electron Zeeman energy to the
nuclear dipole-dipole reservoir is another possible channel for longitudinal
relaxation.~\cite{Abragam61} This, however, can be effective only in the
external magnetic fields comparable or weaker than the local nuclear fields,
which are generally sufficiently small. In addition to those listed in
Table~I, the simultaneous flip-flop of several spins generally entails
energy exchange with the phonon or spin-spin reservoir. Much less attention
has been paid to such processes in the literature.~\cite%
{Al'tshuler74,Sousa031,Cheng04}

\subsection{Elastic processes}

The elastic processes can be imagined as a result of uncontrollable
fluctuations of Zeeman frequency or its equivalent, i.e., the fluctuations
of longitudinal magnetic field that conserve the electron spin projection on
the quantization axis. As summarized in Table~I, there are at least four
different manifestations of this process. Of these, the first one listed
("anharmonic vibration") represents the case that the irregular phonon phase
disturbances originating from the phonon relaxation induces a net effect of
electron spin phase decay via the spin-phonon interaction. If the phonons
are treated as harmonic oscillators without damping, the longitudinal
component of the "effective" interaction field takes the form of a harmonic
perturbation and does not cause alteration in the electron spin phase.~\cite%
{SKdecay04} [One exception may be the small spin phase decoherence ($\sim
10^{-9}$) that can be acquired under the additional assumption of identical
phase for all the harmonic phonon modes at the initial instant.~\cite%
{Mozyrsky02}] However, a different situation can be realized when the phonon
oscillation is interrupted by a series of random phase disturbances as
schematically illustrated in Fig.~3. The reason of such phonon phase steps
can be lattice anharmonicity, phonon scattering at the impurities or lattice
defects, etc. These irregular phonon phase perturbations affect the electron
spin precession resulting in its phase relaxation.~\cite{SKdecay04}

The fluctuations in the local effective magnetic field (and, thus, spin
decoherence) can also result from the transitions between the electronic
(orbital) states with different Zeeman frequencies. As the electron
experiences the stochastic transitions between these orbital states without
spin flip, electron spin phase can change due to the finite and random
lifetime in the states with different precession frequencies (Fig. 4).~\cite%
{SKprl04} It should be emphasized that although the transitions are mediated
by phonons, the energy exchange between the Zeeman and phonon reservoirs
does not take place in contrast to the Orbach process; i.e., this case
undoubtedly belongs to the elastic processes.

Electron spin can accumulate the phase shift through random thermal changes
in the nuclear spin distribution as well. Although a quantum state of
individual nuclear spin generally has a relatively long lifetime, a large
ensemble of nuclear spins (or spin flip transitions) produces a rapidly
evolving effective field distribution. Its significance can be gauged in
terms of electron resonant frequency diffusion (i.e., the so-called \emph{%
spectral diffusion}) in the spin-echo experiments.~\cite%
{Klauder62,Mims68,Chiba72,Sousa03,Tyryshkin03} A peculiar property of this
process is the cubic exponential decay of spin-echo signal $\sim \exp
[-t^{3}/T_{M}^{3}]$, where the memory time $T_{M}$ can be calculated in the
model of uncorrelated nuclear spin flips\cite{Chiba72} or in the model
taking into account the flip-flop processes for nuclear spin pairs.\cite%
{Sousa03} Note that the spectral diffusion process is probably irrelevant to
QC where only the initial phase decay of $\sim 10^{-4}-10^{-6}$ matters for
fault tolerant operation. In the corresponding short time frame, the
processes with a linear exponential decay prevail as demonstrated in Ref.~%
\onlinecite{Tyryshkin03}. Decoherence can also be attributed to isolated
(i.e. non-interacting) nuclear spins.~\cite{Khaetskii02} This is because
each nuclear spin precesses around the electron hyperfine field, which is
inhomogeneous over the QD volume. Such uncorrelated nuclear spin precession,
in turn, leads to the change in the strength of nuclear hyperfine
interaction field affecting the electron spin and its phase. The efficiency
of this process, however, is much weaker than that of spectral diffusion
when the inter-nuclear interaction is taken into account.

The elastic processes in Table~I include the one mediated by fluctuations of
Berry geometric phase.~\cite{Gamliel89,Serebrennikov94,Serebrennikov04} The
efficiency of this process compared to the others is not clearly understood
yet. In addition, an elastic phonon scattering on a group of spins can cause
them out of phase.~\cite{Sousa031,Cheng04} However, this process is less
relevant for single qubit decoherence.

\subsection{Mechanisms of spin-phonon interaction}

Table~II summarizes the microscopic mechanisms of spin-phonon interaction
that give rise to the elastic and inelastic spin relaxation processes
(marked by \# in Table~I). Despite the great multiplicity of the mechanisms
considered in the literature, they can be classified into three groups
following the global characteristics; i.e., the mechanisms resulting from
the (i) spin-orbit, (ii) hyperfine, and (iii) spin-spin interactions. In
turn, these interactions manifest themselves in a number of different ways,
all of which can contribute to spin phase and longitudinal relaxation.

Taking into account that interactions (i), (ii) and (iii) mentioned above
possess, generally speaking, non-zero matrix elements between the states
appurtenant to the Zeeman doublet, their modulation by phonon vibrations can
be expressed as a Hamiltonian realizing the spin-phonon interaction (listed
as "Direct" in Table~II). Mathematically, if a phonon mediates an electronic
potential $V_{ph}(\mathbf{r})$, the spin-orbit interaction is changed by the
term $H_{so-ph}=\frac{\hbar }{2m^{2}c^{2}}[\nabla V_{ph}\times \mathbf{p}]%
\mathbf{s}$ ("SOI modulation" mechanism). At the same time, the shift $%
\Delta \mathbf{R}_{j}$ of an atomic position from the site $\mathbf{R}_{j}$
due to the lattice vibration (i.e., phonons) also changes the Hamiltonian
for the hyperfine interaction $H_{hf}(\mathbf{R}_{j})=\sum_{j}\mathbf{s}%
\widehat{\mathbf{a}}_{hf}(\mathbf{r}-\mathbf{R}_{j})\mathbf{I}$ by $%
H_{hf-ph}=H_{hf}(\mathbf{R}_{j}+\Delta \mathbf{R}_{j})-H_{hf}(\mathbf{R}_{j})
$ ("HFI modulation" mechanism); here, $\widehat{\mathbf{a}}_{hf}(\mathbf{r}-%
\mathbf{R}_{j})$ is the hyperfine interaction tensor. The case of spin-orbit
interaction adapted to the QDs was considered in Ref.~%
\onlinecite{Khaetskii01}. Direct phonon fluctuations of hyperfine
interaction were studied for a donor in silicon,~\cite{Pines57} $F$-centers
in alkali halides,~\cite{Terrile77} and adapted to QDs.~\cite{Abalmassov04}

In a similar manner, one can introduce the spin-phonon interaction mediated
by the spin-spin interactions $H_{ss-ph}$. In this case, one can further
separate the dipole-dipole and exchange interactions. Obviously, these
mechanisms depend on the inter-spin distances. Phonon modulation of
dipole-dipole interaction ("Waller" mechanism) was considered in Ref.~%
\onlinecite{Waller32}; the role of exchange interaction including the
Dzialoshinski-Moria terms~\cite{Dzyaloshinskii58,Moriya60} (that can be
important in crystals with no inversion symmetry~\cite{Kavokin01}) was
analyzed in Ref.~\onlinecite{Al'tshuler74}.

The direct mechanisms considered above are not the only manifestation of the
spin-phonon interactions. In the strict sense, the Kramers doublets actually
represent non-multiplicative functions of spin and space variables due to
the the spin-orbit or hyperfine interactions (i.e., mixing of orbital and
spin states; "Admixture" in Table~II). This makes a spin-independent phonon
potential capable of evoking the transitions between the Kramers doublet.
Such a mechanism becomes apparent in the case of hole spin relaxation in a
QD, where the spin-orbit interaction imposes a linear combination of
different spin and orbital states for the basis of valence band.~\cite%
{Bulaev05,Zinov'eva05} Mathematically, the mechanism constitutes only the
non-zero matrix elements that can be symbolically
transformed to the form
\begin{equation}
\sum_{\left\{ e\right\} }\left[ \frac{\left\langle g\uparrow \right\vert
V_{ph}(\mathbf{r})\left\vert e\uparrow \right\rangle \left\langle e\uparrow
\right\vert H_{mix}\left\vert g\downarrow \right\rangle }{E_{g}-E_{e}}+\frac{%
\left\langle g\uparrow \right\vert H_{mix}\left\vert e\downarrow
\right\rangle \left\langle e\downarrow \right\vert V_{ph}(\mathbf{r}%
)\left\vert g\downarrow \right\rangle }{E_{g}-E_{e}}\right] \,,  \label{p1}
\end{equation}%
if a perturbation theory is applicable for the system with the electron
energies $E_{g}$ and $E_{e}$ in the ground ($\left\vert g\right\rangle $)
and excited ($\left\vert e\right\rangle $) states; $\uparrow $ and $%
\downarrow $ denote different spin states. Equation~(\ref{p1}) assumes that
the diagonal part of the Hamiltonian $H_{mix}$ enters into $E_{g}$ and $E_{e}
$, while the off-diagonal element mixes the ground and excited states with
opposite spins.

In the second-order perturbation theory, Eq.~(\ref{p1}) represents the
spin-phonon interaction as the process of virtual transition to the
spin-flip excited state due to $H_{mix}$ followed by a transition to the
ground state without spin flip via the spin-independent electron-phonon
interaction, or vice versa. In the case of electron in a QD, this mechanism
can possess an advantage over the direct spin-phonon interaction due to the
relatively strong spin coupling between the non-Kramers (i.e., orbital)
states. A distinctive feature of such indirect spin-phonon interactions is
their proportionality to the matrix elements of electron-phonon interaction
between the ground and excited states. Consequently, these mechanisms can be
further classified based on the types of electron-phonon interactions as
well as the potentials that cause effective mixture of the ground and
excited states.


The mechanism caused by simultaneous manifestation of spin-orbit and
spin-independent electron-phonon interactions was initially developed by
Kronig~\cite{Kronig39} and Van Vleck~\cite{VanVleck40} for magnetic ions.
Adaptation of this "Kronig-Van Vleck" mechanism to the conduction electrons
in a QD can be conveniently achieved by extracting the spin-orbit coupling
in the form of the Rashba and Dresselhause terms (which efficiently mix the
QD ground and excited states).~\cite{Bychkov84,D'yakonov86} Then, the phonon
potential of various origins such as the deformation potential,
piezoelectric potential,\cite{Khaetskii01,Golovach04,Fal'ko05,Sherman05} or
the ripple effect at the QD interface\cite{Wood02} gives rise to the
effective spin-phonon coupling in the second order. The spin-orbit coupling
in the valence band is generally much stronger; hence, \emph{holes} are less
attractive as a qubit.\cite{Bulaev05,Zinov'eva05}

When $H_{mix}$ is due to the hyperfine interaction, the spin flip can be
considered in a manner similar to that with the spin-orbit interaction as
discussed in Refs.~\onlinecite{Erlingsson02} and \onlinecite{Abalmassov04};
an isomorphous case of a diluted magnetic semiconductor QD was also
investigated.~\cite{Yang05} It seems that this admixture spin-phonon
interaction is more effective than the direct counterpart (i.e., phonon
modulation of the hyperfine interaction). As to the spin-spin interactions
(both dipole-dipole and exchange), they can also induce the mixing of spin
and orbital states similarly; however, such a treatment has not yet been
developed to the best of our knowledge.

In addition to the direct and admixture mechanisms discussed above, one can
also consider "indirect" manifestation of spin dependent interactions. This
group of mechanisms accounts for the random fluctuations in the effective
field through the phonon-mediated stochastic transitions between different
orbital states. As the characteristics of $g$-tensor, hyperfine field, or
spin-spin interactions vary with the orbital states, electrons experience
uncontrollable changes in these parameters while undergoing spin-independent
phonon scattering between them. Such fluctuations cause the transversal
relaxation of electron spin~\cite{SKprl04} along with a potential
contribution to the longitudinal relaxation. However, the latter (i.e.,
spin-flip) is expected to be small due to the negligibly small influence on
the spectral density at typical Zeeman frequencies.

Finally, there is a higher-order phonon-mediated mechanism that can affect
the spin relaxation. Although the phonon potential $V_{ph}(\mathbf{r})$
cannot influence spin precession directly, its incorporation with the
electron potential $V_{0}(\mathbf{r})$, which contributes to the $g$-tensor
in a different manner,\cite{Kiselev198,Ivchenko97} results in a specific
spin-phonon interaction $H_{g-ph}=\sum (\delta g_{ij}/\delta V_{0})V_{ph}(%
\mathbf{r})\mu _{B}B_{i}s_{j}$, where $s_{j}$ are the components of electron
spin operator. This mechanism was analyzed in a QD with respect to the
Orbach spin-flip processes~\cite{Khaetskii01} as well as in a bulk
semiconductor.~\cite{Margulis83}

In summarizing the possible mechanisms of spin-phonon interaction, it must
be pointed out that all of them make \textit{correlated} contributions to
the processes marked \# in Table I. For example, the direct spin-flip
process results from the phonon-induced fluctuation of the effective
magnetic field (with the resonant frequency $\hbar \omega $) that is
realized by additive contributions from the spin-phonon mechanisms listed in
Table~II (i.e., the spin-orbit, hyperfine, and spin-spin interactions).
Consequently, the corresponding matrix elements of individual mechanisms
must be summed before the probability of any process is calculated. Although
the importance of interference for certain mechanisms was already discussed
in Refs.~\onlinecite{Averkiev99} and \onlinecite{Golovach04}, it requires
further theoretical scrutiny. However, the anticipated difficulty may be
circumvented for reasonably weak magnetic fields if the spin-phonon
Hamiltonian $H_{s-ph}$ is considered in terms of a phenomenological tensor $%
\mathbf{A}$ following the analysis of Ref.~\onlinecite{Koloskova63}. This
parameter $\mathbf{A}$ is designed to represent the net/total contribution
of the spin-orbit mechanisms and can be found experimentally from the
dependence of electron paramagnetic resonance signal on the applied stress.

As for the contributions of different processes (Table I) to spin
relaxation, they do not interfere and, thus, can be considered
independently. As mentioned earlier, the present paper investigate dominant
elastic spin relaxation processes via the spin-phonon mechanisms,
particularly those originated from the spin-orbit and hyperfine interactions.

\section{Basic equation for electron spin evolution}

With the mechanisms responsible for the spin-phonon interaction $H_{SL}$
ascertained in Table~II, the evolution of electron spin can be described
according to the rubrics under "Decoherence ($T_{2}$)" in Table I. In
contrast to the calculation of $H_{SL}$ that requires individualized
treatments for each mechanisms, most processes, elastic and inelastic,
marked by \# in Table I (with an exception of "geometric phase" process) can
be approached by a general formalism based on kinetic equations. For
simplicity, the derivation of these equations is restricted to the first
Born approximation in the present consideration. As such, the two-phonon
processes, which appear on the second order of $H_{SL}$ [Fig. 2(b)], are not
included.

There are several comparable approaches to describe the evolution of spin $%
S=1/2$ interacting with a thermal bath. For the purposes of decoherence
evaluation in the QDs, it is desirable to apply a system of equations
directly for the mean spin components (rather than for the density matrix
components), which analyze their temporal progression on the Bloch sphere
from an initial state to that corresponding to the thermal equilibrium of
the system. At the same time, it is desirable to express the relaxation
coefficients in terms of the spectral density of the thermal bath operators
that allows application of the formalism well-developed in the case of
thermodynamic equilibrium. Finally, the equations must correctly account for
the anisotropy of spin interaction with the thermal bath. Such equations
were studied previously in the cases of specific symmetry properties of the
correlation functions.~\cite{Sem03} This section generalizes and adopts
these equations to the problem of spin decoherence.

It is convenient to represent the Hamiltonian of one electron system in
terms of the effective spin operator $\mathbf{s}$ for the Kramers doublet $%
\{g\}=\{\left\vert g\right\rangle ,\left\vert \widetilde{g}\right\rangle \}$
of the ground electronic state. The linear magnetic splitting of Kramers
doublet $\{g\}$ is described by an effective Hamiltonian in the common form $%
H_{S}=g_{\mu \nu }\mu _{B}B_{\mu }s_{\nu }$, where the $g$-tensor $g_{\mu
\nu }$ is introduced. Similarly, the electron spin interaction $H_{SL}$ with
the thermal bath (of Hamiltonian $H_{L}$) is expressible as a linear form of
the spin operators, $H_{SL}=\mathbf{\Omega }^{(g)}\mathbf{s}$ if only the
states $\{g\}$ are involved in the relaxation. Actually, $H_{SL}$ also
describes the transitions between different Kramers doublets. If these
transitions from the ground $\{g\}$ to excited $\{e\}$ states are virtual
(as is the case at sufficiently low temperatures $kT<<\delta _{0}$, where $%
\delta _{0}$ is the energy separation between the $\{g\}$ and $\{e\}$
states), reduction of the Hamiltonian $H_{SL}$ to the basis $\{g\}$ of spin
operators $\mathbf{s}$ poses no specific problem. For example, the
off-diagonal elements of $H_{SL}$ between $\{g\}$ and $\{e\}$ states can be
eliminated through the procedure of canonical transformation. Hence, the
evolution of the spin system at low temperatures is fully described by an
effective $2\times 2$ Hamiltonian,
\begin{equation}
H=H_{S}+H_{L}+H_{SL}.  \label{f1}
\end{equation}%
This Hamiltonian can also be applicable to the cases other than at low
temperatures. Note that the problems involving a group of Kramers doublets
(e.g., $\{n\}=\{g\}$,$\{e\}$) sometimes allow description by an effective
Hamiltonian $H_{S}^{eff}$ with $\left\langle \{n\}\right\vert
H_{S}^{eff}\left\vert \{n^{\prime }\}\right\rangle =\delta _{n,n^{\prime
}}(E_{n}+g_{\mu \nu }^{(n)}\mu _{B}B_{\mu }s_{\nu })$ and $\left\langle
\{n\}\right\vert H_{SL}\left\vert \{n^{\prime }\}\right\rangle =(\widehat{%
\mathbf{\Omega }})_{n,n^{\prime }}\mathbf{s}$; here, the matrix of orbital
energy $\delta _{n,n^{\prime }}E_{n}$ can alternatively be considered a part
of $H_{L}$ as well. Hence, this case leaves room for the Hamiltonian in the
form of Eq.~(\ref{f1}) even at elevated temperatures. Some examples will be
discussed later.

For a detailed description, we start with the kinetic equation for the spin
density matrix
\begin{equation}
\widehat{\sigma }=\mathrm{Tr}_{L}\rho \,.  \label{f2}
\end{equation}%
Here $\rho $ obeys the Liouville equation $i\overset{\cdot }{\rho }=[H,\rho ]
$, the trace is taken over all variables of the system except the given
electron spin, and $\hbar $ and $k_{B}$ are set to $1$ in this section.
Assuming that the thermal bath is in a thermodynamic equilibrium, one can
introduce its statistical operator
\begin{equation}
f=\frac{\exp \left( -\beta H_{L}\right) }{\mathrm{Tr}_{L}\exp \left( -\beta
H_{L}\right) };~~~\beta =\frac{1}{T}.  \label{f3}
\end{equation}%
Following the approach developed in Ref.~\onlinecite{AK} with a projection
operator method (see Ref.~\onlinecite{Zwanzig60}), the kinetic equation of
operator $\widehat{\sigma }$ in the Born approximation is obtained as
\begin{equation}
\frac{d\widehat{\sigma }\left( t\right) }{dt}=-i[H_{S},\widehat{\sigma }%
(t)]-\int_{0}^{t}\mathrm{\ Tr}_{L}[V\left( t,t\right) ,[V\left( t,t^{\prime
}\right) ,f\widehat{\sigma }\left( t\right) ]]dt^{\prime },  \label{f4}
\end{equation}%
where $[A,B]=AB-BA$. Equation~(\ref{f4}) implies renormalization of the
spin-bath interaction
\begin{equation}
V=\left( \mathbf{\Omega }^{(g)}-\left\langle \mathbf{\Omega }%
^{(g)}\right\rangle \right) \mathbf{s},  \label{f5}
\end{equation}%
so that the averaging over the bath $\left\langle ...\right\rangle =\mathrm{%
Tr}_{L}f...$ results in $\left\langle V\right\rangle =0$ while $%
H_{S}=H_{S}^{(g)}+\left\langle \mathbf{\Omega }^{(g)}\right\rangle \mathbf{s}
$. The last term in Eq.~(\ref{f4}) resembles a collision integral for the
common kinetic equation, where the possible influence of alternating
magnetic fields is ignored. In the case of time-independent $H_{S}$, the
operators on the right-hand side of Eq.~(\ref{f4}) take the forms
\begin{eqnarray}
&&V(t,t^{\prime })=\exp \left\{ -iH_{S}\left( t-t^{\prime }\right) \right\}
V(t^{\prime })\exp \left\{ iH_{S}\left( t-t^{\prime }\right) \right\} ,
\label{f6} \\
&&V(t)=\exp (iH_{L}t)V\exp (-iH_{L}t).  \label{f7}
\end{eqnarray}

Although the integrand in Eq.~(\ref{f4}) can be calculated for arbitrary
coordinates, its subsequent application is not very convenient. Without the
loss of generality, the $z$ axis can be chosen along the quantization
direction that reduces the spin Hamiltonian to
\begin{equation}
H_{S}=\omega s_{z}.  \label{f8}
\end{equation}%
Note that the $z$ axis, generally speaking, does not coincide with the
direction of the external magnetic field if the $g$-tensor reflects the low
symmetry of a QD and $\omega =\sqrt{\sum_{i}\left( g_{i}\mu
_{B}B_{i}+\left\langle \Omega _{i}^{(g)}\right\rangle \right) ^{2}}\equiv
g_{eff}\mu _{B}B$. Here $g_{i}$ are the principal values of $g$-tensor while
$B_{i}$ and $\Omega _{i}^{(g)}$ are the vector components along the main
axes of $g$-tensor.\cite{comm1a} By adopting the specified coordinate
system, Eqs.~(\ref{f6}) and (\ref{f7}) can be easily calculated with $\Omega
_{\mu }=\Omega _{\mu }^{(g)}-\left\langle \Omega _{\mu }^{(g)}\right\rangle $%
,%
\begin{eqnarray}
V(t,t^{\prime }) &=&\Omega _{x}(t^{\prime })(s_{x}\cos u+s_{y}\sin u)+\Omega
_{y}(t^{\prime })(s_{y}\cos u-s_{x}\sin u)+\Omega _{z}(t^{\prime })s_{z},
\label{f10} \\
V(t) &=&\Omega _{x}(t)s_{x}+\Omega _{y}(t)s_{y}+\Omega _{z}(t)s_{z},
\label{f10a}
\end{eqnarray}%
where $u=\omega (t-t^{\prime })$ and $\Omega _{\mu }(t)$ is defined similar
to Eq.(\ref{f7}). One can see that the collision integral can be expressed
in terms of double-time correlation functions
\begin{equation}
\left\langle \Omega _{\mu }(t),\Omega _{\nu }(t^{\prime })\right\rangle
=Tr_{L}\{e^{-\beta H_{L}}\Omega _{\mu }(t),\Omega _{\nu }(t^{\prime })\}
\label{f11}
\end{equation}%
with an evident property $\left\langle \Omega _{\mu }(t),\Omega _{\nu
}(t^{\prime })\right\rangle =\left\langle \Omega _{\mu }(t-t^{\prime
}),\Omega _{\nu }(0)\right\rangle =\left\langle \Omega _{\mu }(0),\Omega
_{\nu }(t^{\prime }-t)\right\rangle $. We consider that these correlation
functions reduce to zero as soon as $\left\vert t-t^{\prime }\right\vert $
exceeds the correlation time $\tau _{c}$. Thus, two regimes of spin
evolution must be distinguished (Ref.~\onlinecite{Sher60}). In the first
regime for short times $t\ll \tau _{c}$, the relaxation coefficients in Eq.~(%
\ref{f4}) are proportional to $t$. This leads to a quadratic-on-$t$
deviation from the initial states of the system that reproduces the very
general results obtained previously in various analyses (see Refs.~%
\onlinecite{Khalfin68} and \onlinecite{Urbanowski94}). Such a non-linearity
results in only a marginal effect of spin transition from the ground state
at sufficiently short times (Zeno effect\cite{Misra77}). Consequently, we
focus on the second, opposite case of reasonably long time $t\gg \tau _{c}$,
when the upper limit of the integral in Eq.~(\ref{f4}) can tend to infinity,
$t\rightarrow \infty $. In this case, the relaxation coefficients are
independent of time and Eq.~(\ref{f4}) describes the Markovian process with
exponential evolution of $\widehat{\sigma }(t)$.

To proceed further, it is convenient to introduce the Fourier transformation
of the correlation functions [Eq.~(\ref{f11})],%
\begin{equation}
\gamma _{\mu \nu }(\omega )\equiv \left\langle \Omega _{\mu }(\tau ),\Omega
_{\nu }(0)\right\rangle _{\omega }=\frac{1}{2\pi }\int_{-\infty }^{\infty
}\left\langle \Omega _{\mu }(\tau ),\Omega _{\nu }(0)\right\rangle
e^{i\omega \tau }d\tau .  \label{f12}
\end{equation}%
Considering the functional form of Eq.~(\ref{f4}), another useful relation is%
\begin{equation}
\int_{0}^{\infty }\left\langle \Omega _{\mu }(\tau ),\Omega _{\nu
}(0)\right\rangle e^{i\omega \tau }d\tau =\pi \gamma _{\mu \nu }(\omega
)+iP\int_{-\infty }^{\infty }\frac{\gamma _{\mu \nu }(\omega ^{\prime })}{%
\omega -\omega ^{\prime }}d\omega ^{\prime },  \label{f13}
\end{equation}%
where we use the symbolic identity $\int_{0}^{\theta }e^{i\omega \tau }d\tau
=\pi \delta (\omega )+iP\frac{1}{\omega }$; $P$ denotes the principal value
of the integral. Finally, let us express the density matrix $\widehat{\sigma
}(t)$ in terms of mean values $\left\langle s_{x}\right\rangle \equiv
\mathrm{X}$, $\left\langle s_{y}\right\rangle \equiv \mathrm{Y}$, and $%
\left\langle s_{z}\right\rangle \equiv \mathrm{Z}$,%
\begin{equation}
\widehat{\sigma }(t)=\frac{1}{2}\left[ \widehat{1}+4\left( \mathrm{{X}s_{x}+{%
Y}s_{y}+{Z}s_{z}}\right) \right] ,  \label{f14}
\end{equation}%
so that $Tr_{s}\widehat{\sigma }(t)s_{\mu }=\left\langle s_{\mu
}\right\rangle $, where $\widehat{\sigma }(t)$ is taken from Eq. (\ref{f14}%
); $\widehat{1}$ is the $2\times 2$ identity matrix and the trace $Tr_{s}$
is taken over spin states. Multiplying Eq.~(\ref{f4}) sequentially by $s_{x}$%
, $s_{y}$ and $s_{z}$ from the right and taking $Tr_{s}$, one can obtain
kinetic equations for $\mathbf{\mathbf{S}}\mathrm{{=\{X},{Y,Z\}}}$, which
define the spin vector evolution on the Bloch sphere in the most general
form
\begin{equation}
\frac{d}{dt}\mathbf{S}=\overrightarrow{{\mathbf{\omega }}}\times \mathbf{S}-{%
\Gamma }\left( \mathbf{S}-\mathbf{S}_{0}\right) \,.  \label{f14a}
\end{equation}%
In this equation, the $3\times 3$ matrix ${\Gamma }$ represents a rather
complicated expression, $\overrightarrow{\mathbf{\omega }}=\{0,0,\omega \}$,
$\mathbf{\mathbf{S}}_{0}=\{0,0,\mathrm{{Z}_{0}\}}$, and $\mathrm{{Z}_{0}=-%
\frac{1}{2}\tanh \frac{\beta \omega }{2}}$ is a static spin polarization due
to the spin splitting $\omega $.

The matrix ${\Gamma }$ can be simplified notably if one takes into account
the symmetry properties of the Fourier transformation of correlation
functions [Eq.~(\ref{f12})],\cite{comm1}%
\begin{equation}
\left\langle \Omega _{\mu }(\tau ),\Omega _{\nu }\right\rangle _{\omega
}=e^{\beta \omega }\left\langle \Omega _{\nu }(\tau ),\Omega _{\mu
}\right\rangle _{-\omega }=e^{\beta \omega }\left\langle \Omega _{\nu
},\Omega _{\mu }(\tau )\right\rangle _{\omega }.  \label{f15}
\end{equation}%
At the same time, only the symmetrical part $\overline{\Gamma }=\left\Vert
(\Gamma _{\mu \nu }+\Gamma _{\nu \mu })/2\right\Vert $ is relevant to spin
relaxation. This is because the asymmetrical part $\Delta \Gamma $ (${\Gamma
}=\overline{\Gamma }+\Delta {\Gamma }$) contributes to Eq.~(\ref{f14a}) as
an additional effective magnetic field that cannot reduce the length of
vector $\Delta \mathbf{S}=\mathbf{S}-\mathbf{S}_{0}$: $d\Delta \mathbf{S/}dt=%
\frac{\Delta \mathbf{S}}{\Delta \mathrm{S}}\cdot \frac{d\Delta \mathbf{S}}{dt%
}=\frac{\Delta \mathbf{S}}{\Delta \mathrm{S}}\cdot \lbrack (\overrightarrow{%
\mathbf{\omega }}-$ $\overrightarrow{\Delta \omega })\times \Delta \mathbf{S}%
]-\frac{\Delta \mathbf{S}}{\Delta \mathrm{S}}\cdot \overline{\Gamma }\Delta
\mathbf{S}=-\frac{\Delta \mathbf{S}}{\Delta \mathrm{S}}\overline{\Gamma }%
\Delta \mathbf{S}$; $\Delta \omega _{x}=-\Delta \Gamma _{yz}$, etc.
Subsequently, the contribution of $\Delta \Gamma $ to spin relaxation is
ignored. We also neglect in the calculation the second term on the
right-hand side of Eq.~(\ref{f13}) (proportional to $iP$) since it plays an
insignificant role as well.~\cite{AK,Slichter,Zevin75} The final result of
the symbolic computation for the matrix ${\Gamma }$ reads
\begin{equation}
\overline{{\Gamma }}=\pi \left(
\begin{array}{ccc}
\gamma _{zz}^{0}+n(\overline{\gamma }_{yy}+\widetilde{\gamma }_{xy}) & -n%
\overline{\gamma }_{xy} & -\frac{n}{2}(\overline{\gamma }_{xz}-\widetilde{%
\gamma }_{yz})-\frac{1}{2}\gamma _{xz}^{0} \\
-n\overline{\gamma }_{xy} & \gamma _{zz}^{0}+n(\overline{\gamma }_{xx}+%
\widetilde{\gamma }_{xy}) & -\frac{n}{2}(\overline{\gamma }_{yz}-\widetilde{%
\gamma }_{zx})-\frac{1}{2}\gamma _{yz}^{0} \\
-\frac{n}{2}(\overline{\gamma }_{xz}-\widetilde{\gamma }_{yz})-\frac{1}{2}%
\gamma _{xz}^{0} & -\frac{n}{2}(\overline{\gamma }_{yz}-\widetilde{\gamma }%
_{zx})-\frac{1}{2}\gamma _{yz}^{0} & n(\overline{\gamma }_{xx}+\overline{%
\gamma }_{yy})+2n\widetilde{\gamma }_{xy}%
\end{array}%
\right) ,  \label{f16}
\end{equation}%
where $n=n(\omega )=(1+e^{-\beta \omega })/2$, $\overline{\gamma }_{\mu \nu
}=[\gamma _{\mu \nu }(\omega )+\gamma _{\nu \mu }(\omega )]/2$, $\widetilde{%
\gamma }_{\mu \nu }=i[\gamma _{\mu \nu }(\omega )-\gamma _{\nu \mu }(\omega
)]/2$, and $\gamma _{\mu ,\nu }^{0}=\gamma _{\mu \nu }(0)$.\cite{comm2}

If the correlation functions possess some supplementary symmetry, Eq.~(\ref%
{f16}) allows further simplification. Assuming $\gamma _{\mu \nu }(\omega
)=\gamma _{\nu \mu }(\omega )$, one can find%
\begin{equation}
\overline{ {\Gamma }}=\pi \left(
\begin{array}{ccc}
\gamma _{zz}^{0}+n\overline{\gamma }_{yy} & -n\overline{\gamma }_{xy} & -%
\frac{n}{2}\overline{\gamma }_{xz}-\frac{1}{2}\gamma _{xz}^{0} \\
-n\overline{\gamma }_{xy} & \gamma _{zz}^{0}+n\overline{\gamma }_{xx} & -%
\frac{n}{2}\overline{\gamma }_{yz}-\frac{1}{2}\gamma _{yz}^{0} \\
-\frac{n}{2}\overline{\gamma }_{xz}-\frac{1}{2}\gamma _{xz}^{0} & -\frac{n}{2%
}\overline{\gamma }_{yz}-\frac{1}{2}\gamma _{yz}^{0} & n(\overline{\gamma }%
_{xx}+\overline{\gamma }_{yy})%
\end{array}%
\right) ,  \label{f17}
\end{equation}%
which readily transforms to the relaxation matrix obtained in Ref.~%
\onlinecite{Pikus84} in the limit $\omega \rightarrow 0$. In other case with
$\gamma _{\mu \nu }(\omega )=-\gamma _{\nu \mu }(\omega )$ for $\mu \neq \nu
$, Eq.~(\ref{f16}) can be put in a simpler form
\begin{equation}
\overline{ {\Gamma }}=\pi \left(
\begin{array}{ccc}
\gamma _{zz}^{0}+n(\overline{\gamma }_{yy}+\widetilde{\gamma }_{xy}) & 0 &
\frac{n}{2}\widetilde{\gamma }_{yz} \\
0 & \gamma _{zz}^{0}+n(\overline{\gamma }_{xx}+\widetilde{\gamma }_{xy}) &
\frac{n}{2}\widetilde{\gamma }_{zx} \\
\frac{n}{2}\widetilde{\gamma }_{yz} & \frac{n}{2}\widetilde{\gamma }_{zx} &
n(\overline{\gamma }_{xx}+\overline{\gamma }_{yy})+2n\widetilde{\gamma }_{xy}%
\end{array}%
\right) .  \label{f18}
\end{equation}
Note that Eqs.~(\ref{f14a}) and (\ref{f16}) are, generally speaking, derived
for the case of anisotropic medium and arbitrary electron spin splitting $%
\omega $. Equation~(\ref{f16}) can be reduced to the common Bloch equations
if one puts $T_{1}^{-1}=\pi n\left[ \gamma _{xx}(\omega )+\gamma
_{yy}(\omega )\right] $, $T_{2}^{-1}=\pi \gamma _{zz}(0)+T_{1}^{-1}/2$ and
omits the off-diagonal components of $\Gamma $. However this reduction does
not always happen, as it was shown in Ref.~\onlinecite{Sem03}. The spectral
properties of Eq.~(\ref{f15}) with respect to the definition of $n(\omega ) $
reveal that the relaxation matrix given in Eq.~(\ref{f16}) is an even
function of electron spin splitting (i.e., ${\Gamma }(\omega )= {\Gamma }%
(-\omega )$), as it also follows from the time inversion symmetry.\cite%
{Landau}

The kinetic equations described above do not assume any specific properties
of the heat bath ($H_{L}$) nor the interaction mechanism ($H_{SL}$), and
form the backbone of our theoretical approach. In the following, we show how
they can be applied to the problem of QD spin decoherence, particularly the
elastic processes described by $\gamma _{zz}(0)$ in Eq.~(\ref{f16})
corresponding to "anharmonic vibration" and "local field steps" in Table I.

\section{Anharmonic vibration}

There are several manifestations of the elastic relaxation process termed
"anharmonic vibration". To avoid the difficulties in determining adequate
mechanisms of spin-phonon interaction and their interference, we consider
this problem semi-phenomenologically by applying the parameters of the
spin-deformation Hamiltonian,~\cite{Koloskova63} which can be extracted from
the electron paramagnetic resonance (EPR) experiments on shallow donors in
strained crystals or known dependencies of electron $g$-tensor shift in a
low-symmetry potential. These parameters are mainly associated with a direct
modulation of spin-orbit interaction as discussed above (see Table~II). We
also take into account the "HFI modulation" mechanism when the
spin-deformation Hamiltonian is not applicable. A brief preliminary account
of this process was given earlier.~\cite{SKdecay04}

\subsection{Qualitative analysis}

Our analysis is based on the representation of spin-phonon interaction in
terms of fluctuating effective magnetic field $\mathbf{\Omega }$\ [in units
of energy, see Eq.~(\ref{f5})] acting on the electron spin $\mathbf{s}$.
This field is assumed to be composed of additive contributions $\mathbf{\
\Omega }_{p}$ from each phonon $p=\{\mathbf{q},\varkappa \}$ with a wave
vector $\mathbf{q}$ and polarization $\varkappa $, i.e., $\mathbf{\Omega }
=\Sigma _{p}\mathbf{\Omega }_{p}$. For the moment, let us focus on a single
phonon contribution. Then, in the frame of reference rotating with a Zeeman
frequency, the electron spin performs precession around the small $\mathbf{\
\Omega }_{p}$, which oscillates with the phonon frequency $\omega _{p}$. No
alteration in the electron spin phase occurs due to such a harmonic
perturbation with a possible exception of spin phase shift $\Delta \phi _{0}$
acquired at the initial period of interaction $0<t<2\pi /\omega _{p}$ due to
a random phonon phase $\theta _{p}$.~\cite{Klemens}

A different situation can be realized when the phonon harmonic oscillation
is interrupted and resumes at a series of instant times $t_{1i}$ and $t_{2i}$
($i=1,2,...$), respectively, as shown in Fig.~3. The reason of such phonon
fluctuations can be lattice anharmonicity, phonon scattering at the
impurities or lattice defects, etc. These irregular phonon perturbations
affect the electron spin precession resulting in the phase shift $\Delta
\phi _{i}$ at each interval of time $t_{2i}-t_{1i}$. Subsequently, the net
effect of spin phase change $\phi _{p}(t)$ due to a phonon mode $p$ can be
expressed as $\phi _{p}(t)=\Sigma _{i}\Delta \phi _{i}$, ($t_{2i}<t$).

Note that for a large number of small changes $\Delta \phi _{i}$, their
total effect can be described by a diffusion equation. Its solution leads to
an exponential decay of electron spin phase with a relaxation rate $%
T_{p}^{-1}=\frac{1}{2}\left\langle \Delta \phi _{i}^{2}\right\rangle \tau
_{p}^{-1}$, where $\tau _{p}$ is the mean time between sequential instants $%
t_{1i}$ (or $t_{2i}$).\cite{Sem03} To estimate the spin phase change $\Delta
\phi _{i}$ caused by a phonon perturbation during $t_{2i}-t_{1i}$, it is
helpful to recognize that the oscillator does not change the spin phase
during its full period $\Delta t_{p}=2\pi /\omega _{p}$ as well as for any $%
n $ integer periods $n2\pi /\omega _{p}$. Hence, $\Delta \phi _{i} $ can be
approximated as a spin rotation $\Omega _{p}\Delta t_{p}$ in an effective
field $\mathbf{\Omega }_{p}$ independently of the duration $t_{2i}-t_{1i} $.
With the mean value $\left\langle \Delta \phi _{i}\right\rangle $ on the
order of $\Omega_{p}/ \omega_{p}$, one can expect $T_{p}^{-1} \sim
\tau_{p}^{-1} \Omega_{p}^{2}/\omega _{p}^{2}$ for the phonon mode $p$ and $%
T_{2}^{-1}\sim \Sigma_{p} N_{p} \tau_{p}^{-1} \Omega_{p}^{2}/\omega _{p}^{2}$
when the contributions of all phonons (with the population factor $N_{p}$)
are taken into account.

\subsection{Theoretical model}

The qualitative consideration provided above shows that electron spin phase
relaxation can be strongly affected by the phonon phase damping of any
origin such as phonon decay. Since this mechanism does not involve energy
exchange, only the longitudinal (with respect to the external magnetic field
$\mathbf{B}$) component $\Omega _{z}$ of the effective fluctuating field is
relevant to our case. Since the random single spin fluctuation associated
with each phonon scattering is expected to be very small, we investigate the
total result of these small fluctuations averaged over the time scale $%
\Delta t$ ($\tau _{c}\ll \Delta t\ll T_{2}$). Thus, the time evolution of
the mean spin value $\mathbf{s}$ can be described in terms of the quantum
kinetic equation discussed in the previous section~[Eqs.~(\ref{f14a}) and (%
\ref{f17})].\cite{Sem03} The analysis of $T_{2,0}^{-1}=\pi \gamma _{zz}^{0}$
is the focus of the investigation as it determines the transversal
relaxation stemming from the elastic processes.

The $z$ component of the fluctuating field operator due to the phonon decay
can be expressed in a form linear in the creation and annihilation operators
$a_{p}^{\dag }$ and $a_{p}$ of the phonon mode $p=\{\mathbf{q},\varkappa \}$
($\mathbf{q}$ and $\varkappa $ are the phonon wave vector and its
polarization, $-p\equiv \{-\mathbf{q},\varkappa \}$); i.e.,
\begin{eqnarray}
\Omega _{z} &=&\sum_{p}V_{z}^{p}Q_{p};  \label{r5} \\
Q_{p} &=&a_{p}^{\dag }-a_{-p},  \label{r6}
\end{eqnarray}%
with a matrix element $V_{z}^{p}$ of the spin-phonon interaction, whose
specific form will be discussed later. Skipping the detailed elaboration of
the routine calculation of the phonon correlation functions,\cite%
{Zubarev60,Pathak65} the final result for the spin relaxation rate
associated with the elastic "anharmonic vibration" process is obtained as
\begin{equation}
T_{2,0}^{-1}=\sum_{p}\left\vert V_{z}^{p}\right\vert ^{2}\left(
2N_{p}+1\right) \frac{\tau _{p}^{-1}}{\omega _{p}^{2}},  \label{eq5}
\end{equation}%
where $\tau _{p}$ is the phonon relaxation time (see Ref.~%
\onlinecite{Klemens93} and the references therein) and the phonon population
factor $N_{p}$ is given as $\left[ \exp \left( \hbar \omega
_{p}/k_{B}T\right) -1\right] ^{-1}$. This expression is in accordance with
the qualitative analysis discussed in previous section with the exception of
the second term in the factor $2N_{p}+1$ (i.e., "1") that is attributed to
the zero-point oscillations. It reflects the quantum effects, which could
not be described in terms of classical approach but may be important at low
temperatures.

To evaluate Eq.~(\ref{eq5}), one needs the knowledge of the relevant phonon
mode $p$ (such as the dispersion relation $\omega _{p}$ and the relaxation
time $\tau _{p}$) as well as its interaction with electron spins. Taking
into account the conditions frequently adopted in QC, we assume that the
radius of the electronic state $a_{0}$ (i.e., the QD size) is much larger
than the lattice constant. This restricts the summation in Eq.~(\ref{eq5})
to long wavelength phonons as the matrix element $V_{z}^{p}$ falls off
quickly for $q>\pi /a_{0}$. Accordingly, $\tau _{p}=\tau _{\varkappa }(%
\mathbf{q})$, which is a complex function of the temperature and phonon
frequencies [see Eq.~(6.2) in Ref.~\onlinecite{Carruthers61}], can be
considered in the long wavelength limit. Of a number of potential phonon
decay mechanisms, only one term originating from the boundary scattering
survives at sufficiently low temperatures $T\lesssim T_{bs}$ ($T_{bs}\approx
10$ K in the case of Ref.~\onlinecite{Carruthers61}). Since this mechanism
is insensitive to the temperature,~\cite{Klemens93} it is adequate to assume
a constant phonon relaxation time $\tau _{p}\simeq \tau _{ph}$. The simple
relaxation time approximation circumvents the difficult problem of the
zone-edge phonons and the dependence $\tau _{p}=\tau _{\varkappa }(\mathbf{q}%
)$, which could be very specific for each particular sample.


\subsection{Effect of acoustic modes}

To evaluate Eq.~(\ref{eq5}) (i.e., $\left\vert V_{z}^{p}\right\vert ^{2}$)
for the acoustic phonons, we consider a phenomenological Hamiltonian of the
spin-lattice interaction. It can be expressed in terms only of symmetry and
does not depend on the specific mechanism;~\cite{Koloskova63}
\begin{equation}
H_{SL}^{ac}=\sum_{ijkl}A_{ijkl}\mu _{B}B_{i}s_{j}\overline{u_{kl}}%
+\sum_{ijklm}A_{ijklm}^{\prime }\mu _{B}B_{i}s_{j}\overline{w_{klm}}\,.
\label{r7}
\end{equation}%
Here $\overline{u_{kl}}$ is the strain tensor $u_{kl}=(\partial \Delta
x_{k}/\partial x_{l}+\partial \Delta x_{l}/\partial x_{k})/2$ averaged over
the electron ground state $\left\vert g\right\rangle $ (i.e., $\overline{%
u_{kl}}=\left\langle g\left\vert u_{kl}\right\vert g\right\rangle $),
subscripts $ijkl$ refer to the crystallographic axes, and the non-zero
components of the tensor $A_{ijkl}$ can be found from the symmetry of the QD.%
\cite{Koloskova63,GK03} The second term in Eq.~(\ref{r7}) takes into account
the effect of strain gradient $w_{klm}=\partial u_{kl}/\partial x_{m}$,
which will be neglected from consideration since it can be irrelevant to the
elastic spin-phonon processes.~\cite{Golovach04}

As a specific problem, we consider a lateral [001] QD with a $z$-directed
magnetic field and a localized electron with the axial symmetry with respect
to the $z$ axis. This case can be applied to high-symmetry systems
corresponding, for example, to shallow donors, etc. Now Eq.~(\ref{r7}) can
be reduced to the form $H_{SL}^{ac}=$ $\mathbf{\Omega }\mathbf{s}$ with $%
\Omega _{z}=\left[ \left( A_{33}-A_{31}\right) \overline{u_{zz}}+A_{31}%
\overline{\Delta }\right] \mu B$ and $\Omega _{x}=\Omega _{y}=0$.\cite{GK03}
$\Delta $ denotes the dilatation $\Delta =\Sigma _{i}u_{ii}$ and the Voigt
notation is adopted ($A_{33}=A_{zzzz}$, $A_{31}=A_{zzxx}$, $A_{66}=A_{xyxy}$%
). Then, as $\Omega _{z}$ is expressed in terms of the creation and
annihilation operators [Eq.~(\ref{r5})], the matrix element $V_{z}^{p}$ is
given as
\begin{equation}
V_{z}^{p}=i\left( \frac{\hbar }{2\rho V_{0}\omega _{p}}\right) ^{1/2}\left[
\left( A_{33}-A_{31}\right) e_{z}^{p}q_{z}+\delta _{\varkappa ,L}A_{31}q%
\right] \Phi \left( \mathbf{q}\right) \mu _{B}B \,.  \label{r8}
\end{equation}%
Here $\rho $ is the mass density of the crystal, $q=\sqrt{%
q_{x}^{2}+q_{y}^{2}+q_{z}^{2}}$, $V_{0}$ is the volume of the sample
structure, $\mathbf{e}^{p}$ the polarization vector of the phonon $p$, and $%
\Phi \left( \mathbf{q}\right) =\left\langle g\right\vert e^{i\mathbf{q}\cdot
\mathbf{r}}\left\vert g\right\rangle $.

The spin-lattice relaxation rate in Eq.~(\ref{eq5}) can be calculated by
treating the acoustic phonon modes based on the isotropic elastic continuum
model with the longitudinal and transverse sound velocities $v_{L}$ and $%
v_{T}$. Then by taking advantage of the phonon relaxation time approximation
(in the long wavelength limit and at sufficiently low temperatures) as well
as the axial symmetry for the localized electron [i.e., $\Phi \left( \mathbf{%
q}\right) =\Phi \left( x,z\right) $ with $x=qa_{0}/2$, $z=q_{z}/q$], one can
obtain
\begin{equation}
T_{2,0}^{-1}=\tau _{ph}^{-1}\xi (B)\displaystyle\int_{0}^{\infty
}x\sum_{\varkappa =L,T}\frac{v_{T}^{3}}{v_{\varkappa }^{3}}\coth \left(
\frac{T_{\varkappa }^{eff}}{T}x\right) F_{\varkappa }\left( x\right) dx,
\label{eq9}
\end{equation}%
\begin{eqnarray}
\xi (B) &=&\frac{\left( A_{33}-A_{31}\right) ^{2}\mu _{B}^{2}B^{2}}{2\pi
^{2}\hbar \rho v_{T}^{3}a_{0}^{2}},  \notag \\
F_{\varkappa }\left( x\right)  &=&\displaystyle\int_{-1}^{1}P_{\varkappa
}(z)\left( z^{2}+\zeta \right) ^{2}\Phi ^{2}\left( x,z\right) dz\,,  \notag
\end{eqnarray}%
where we introduce $1/a_{0}^{2}$ and $a_{0}^{2}$ into $\xi (B)$ and
integrant for convenience. In this equation, $P_{L}(z)=\left( z^{2}+\zeta
\right) ^{2}$; $P_{T}(z)=z^{2}(1-z^{2})$, $\tau _{ph}^{-1}$ is the average
phonon relaxation rate as mentioned above, $T_{\varkappa }^{eff}=\hbar
v_{\varkappa }/k_{B}a_{0}$ is the effective temperature, and $\zeta \equiv
A_{31}/\left( A_{33}-A_{31}\right) =-1/3$ if one assumes that the strain
induced part of the effective $g$-tensor $\widetilde{g}_{ij}=\Sigma
_{k,l}A_{ijkl}u_{kl}$ is characterized by zero trace, i.e., $A_{33}+2A_{31}=0
$.

\subsubsection{Spin decoherence of shallow donor}

Let us evaluate spin relaxation of a shallow donor electron with an
effective Bohr radius $a_{B}$ (=$a_{0}$) and $\Phi \left( x,z\right) =\left(
1+x^{2}\right) ^{-2}$. The last factor determines the relevant phonons,
which must be long wavelength modes with $q\lesssim 1/a_{B}$. Then, the
integral in Eq.~(\ref{eq9}) has a simple analytical approximation,
\begin{equation}
T_{2,0}^{-1}=\frac{2\xi (B)\tau _{ph}^{-1}}{45}\left( \sqrt{1+\frac{T^{2}}{%
T_{T}^{2}}}+\frac{2v_{T}^{3}}{3v_{L}^{3}}\sqrt{1+\frac{T^{2}}{T_{L}^{2}}}%
\right) ,  \label{eq11}
\end{equation}%
where $T_{L(T)}=\left( 16/15\pi \right) T_{L(T)}^{eff}$; this result is for
the case of $\mathbf{B}\parallel \lbrack 001]$.

As an example, we consider a Phosphorus shallow donor in Si with $a_{B}=1.8$%
~nm. The phonon relaxation time can be extracted from the low temperature
measurements of Si thermal resistivity~\cite{GlassSlack} in terms of the
theory developed in Refs.~\onlinecite{Callaway} and \onlinecite{Klemens93} ($%
\tau _{ph}=2.4\times 10^{-8}$ s). The spin-phonon coupling constants were
estimated in the works of Refs.~\onlinecite{Roth} and \onlinecite{Hasegawa}.
However, we believe that direct determination of coupling constants by means
of EPR measurements of Si:P under an applied stress gives more reliable
data. A corresponding experiment was performed in Ref.~\onlinecite{Wilson61}%
, where the constant $A_{66}=0.44$ was found. Similarly, our estimation~\cite%
{SKdecay04} obtains $A_{33}=0.31$ and $A_{31}=-0.155$ that gives $%
T_{2,0}^{-1}=1.3\times 10^{-4}$ s$^{-1}$ at the magnetic field of 1 T and
low temperatures $T\ll T_{L(T)}\simeq 10$ K. In another important case of a
Si shallow donor in Al$_{0.4}$Ga$_{0.6}$As, the data on EPR under a uniaxial
stress\cite{Glaser91} provide rather strong spin-phonon constants of $%
A_{33}=19.6$ and $A_{31}=-9.8$. This gives the estimation $%
T_{2,0}^{-1}=6.1\times 10^{-2}$ s$^{-1}$ and $6.1\times 10^{-4}$ s$^{-1}$
for the magnetic fields of 1 T and 0.1 T, respectively, at $T=4$~K under the
assumption that phonon lifetimes are identical in these crystals.

\subsubsection{Spin decoherence in a QD}

Corresponding calculations can be performed for an electron localized in a
QD of $L_{xy}=2a_{0}$ in the lateral width and $L_{z}=\epsilon L_{xy}$ in
the thickness. Under the condition $\epsilon \lesssim 0.1$, an approximate
formula takes the form
\begin{equation}
T_{2,0}^{-1}=\xi (B)\tau _{ph}^{-1}\left( \sum_{i=L,T}b_{i}\sqrt{%
v_{i}^{2}+d_{i}^{2}\frac{T^{2}}{T_{i}^{2}}}\right) ,  \label{eq12}
\end{equation}%
where the fitting coefficients are $b_{T}=1$, $b_{L}=v_{T}^{3}/v_{L}^{3}$, $%
v_{T}=0.33-1.27\epsilon ^{2}$, $d_{T}=0.35-0.395\epsilon ^{2}$, $%
v_{L}=0.97-28.5\epsilon ^{2}$, and $d_{L}=0.40-3.76\epsilon ^{2}$.

To examine the importance of spin phase relaxation caused by the phonon
decay, the result of Eq.~(\ref{eq12}) is compared with the admixture
spin-flip process~\cite{Khaetskii01} in a GaAs QD with $L_{z}=3$~nm and $%
L_{xy}=25$~nm, assuming $\tau _{ph}=2.4\times 10^{-8}$ s and $A_{33}=19.6$.
For the relatively strong magnetic field of 1 T and $T=4$~K, the present
elastic process and the spin-flip give $T_{2,0}^{-1}\approx 0.1$~s$^{-1}$
and $\frac{1}{2}T_{1}^{-1}=T_{2,\omega }^{-1}=10$~s$^{-1}$, respectively,
while for $B=0.1$~T both predict almost the same rate of $\approx 10^{-3}$~s$%
^{-1}$. At lower magnetic fields, the elastic spin relaxation prevails. The
Si QD gives a qualitatively similar (but smaller) result; its numerical
estimation will be discussed in Sec.~VII.

\subsection{Role of optical phonons}

The optical phonons possess high energies compared to the typical energy
scale (such as the thermal energy $k_B T$) and, thus, are generally ignored
in the problem of QC based on QD electron spins.~\cite{Kochelaev60,Huang67}
However, this is not the case in the elastic process under consideration
since the uncontrolled variation of spin phase can happen without the
presence of thermal phonons as shown in Eq.~(\ref{eq5}) at $N_{p}\rightarrow
0$. Zero-point optical vibrations, which survive even at very low
temperatures $T\ll \hbar \omega _{op}$ ($\omega _{op}$ is the optical phonon
frequency), contribute to spin decoherence as it happens with acoustic modes
in the case $T\ll T_{L,T}$ [Eq.~(\ref{eq12})]. The potential significance of
the optical modes stems from their relatively large contributions at small
wave vectors $q$ ($\left\vert V_{z}^{p}\right\vert ^{2}\rightarrow {constant}
$ as $q\rightarrow 0$)~\cite{Bhatt94} as well as the substantially shorter
phonon lifetime. In contrast, $\left\vert V_{z}^{p}\right\vert
^{2}\rightarrow q$ for acoustic phonons in the long wavelength limit.

For a detailed analysis, consider a QD made of a polar crystal. The
longitudinal-optical phonons induce the electron potential
\begin{equation}
\varphi _{op}=\sqrt{\frac{2\pi \hbar \omega _{op}}{V_{0}\widetilde{\kappa }}}%
\sum_{\mathbf{q}}\frac{\exp (i\mathbf{q}\mathbf{r})}{q}(a_{\mathbf{q}}^{\dag
}-a_{-\mathbf{q}}),  \label{o1}
\end{equation}%
where $1/\widetilde{\kappa }=1/\kappa _{\infty }-1/\kappa _{0}$; $\kappa
_{\infty }$ and $\kappa _{0}$ are the corresponding dielectric constants.
Assume that the QD possesses an asymmetrical potential in the growth
direction $z$ due to an effective electric field $\mathbf{F}$ imposed along
this axis (e.g., the gate bias). Since $F$ influences the electron $g$%
-factor,\cite{Ivchenko97} its superposition with $\varphi _{op}$ results in
a spin-phonon interaction $H_{SL}^{op}=\mu _{B}B(dg/dF)d\varphi _{op}/dz$
that is linear in the phonon operator. After substitution of Eq.~(\ref{o1}),
this expression reduces to the form of Eq.~(\ref{r5}) with%
\begin{equation}
\left\vert V_{z}^{p}\right\vert ^{2}=\left( \mu _{B}B\frac{dg}{dF}\right)
^{2}\frac{2\pi \hbar \omega _{op}}{V_{0}\widetilde{\kappa }}\frac{q_{z}^{2}}{%
q^{2}}\Phi ^{2}\left( \mathbf{q}\right) .  \label{o2}
\end{equation}%
Finally, the spin relaxation rate is obtained as:
\begin{eqnarray}
T_{2,0}^{-1} &=&\tau _{op}^{-1}\xi _{op}(B),  \label{o3} \\
\xi _{op}(B) &=&\frac{2}{\sqrt{\pi }\widetilde{\kappa }}\frac{\left( \frac{dg%
}{dF}\mu _{B}B\right) ^{2}}{\hbar \omega _{op}V_{QD}},  \label{o4}
\end{eqnarray}%
where $V_{QD}$ is the QD volume ($=L_{xy}^{2}L_{z}$).

We perform numerical estimation for a GaAs QD of $L_{xy}=50$~nm and $L_{w}=5$%
~nm. Here, evaluating the QD potential asymmetry \textit{a propri} provides
the most significant challenge. Following the calculations performed for Ga$%
_{0.47}$In$_{0.53}$As/InP quantum wells,\cite{Ivchenko97} a typical value
for $\frac{dg}{dF}$ seems to be approximately 0.085~nm/mV  at $F>10$~mV/nm
with $\mathbf{B}$ directed along the growth axis, while for the
perpendicular direction $\frac{dg}{dF}\leq $0.035~nm/mV. As the effect of
optical phonons quenches in the case of a symmetrical potential ($\frac{dg}{%
dF}=0$ with $F=0$), only the upper limit of this contribution can be
estimated, i.e., $\xi _{op}(1\mathrm{T})\leq 7\times 10^{-6}$. Taking into
account that $\tau _{op}^{-1}\simeq 10^{11}$~s$^{-1}$,~\cite{Bhatt94} one
can find $T_{2,0}^{-1}\leq C_{op}B^{2}$ where $C_{op}=7\times 10^{5}$~s$^{-1}
$T$^{-2}$. This estimation shows that the transversal relaxation via
zero-point optical oscillations can be a very effective decoherence
mechanism in asymmetrical structures. 

\subsection{Effects of hyperfine interaction and two-phonon processes}

Let us consider the elastic one-phonon process associated with the
modulation of hyperfine interaction as well as the two-phonon process. Since
these processes result in sufficiently slow/long spin relaxation, we shall
restrict our consideration to a brief discussion.

The Hamiltonian of hyperfine interaction with the nuclear spin $\mathbf{I}%
_{j}$ located at site $j$ with the position $\mathbf{r}_{j}$ takes the form
\begin{equation}
H_{hf}=a_{hf}\sum_{j}\left\vert \psi _{g}(\mathbf{r}_{j})\right\vert ^{2}%
\mathbf{I}_{j}\mathbf{s},  \label{eq13}
\end{equation}%
where $\psi _{g}(\mathbf{r})=\left\vert g\right\rangle $ is an electron wave
function. The spin-phonon interaction due to small oscillations at $\mathbf{r%
}_{j}$ can be expressed in terms of the dilatation operator $\overline{%
\Delta }=\left\langle g\right\vert \Delta \left\vert g\right\rangle $ if the
mean internuclear distance ($\approx n_{I}^{-1/3}$, $n_{I}$ is the nuclear
spin concentration) is shorter than the wavelength of relevant phonons;
i.e., $H_{SL}^{hf}=\mathbf{\Omega }\mathbf{s}$, where $\mathbf{\Omega }=%
\widehat{n}a_{hf}\overline{\Delta } \sqrt{I(I+1)n_{I}/V_{QD}^{hf}}$. Here,
the unit vector $\widehat{n}$ is directed along the effective nuclear field
defined by Eq.~(\ref{eq13}) and $V_{QD}^{hf}=\left( \int \left\vert \psi (%
\mathbf{r})\right\vert ^{4}d^{3}\mathbf{r}\right) ^{-1}$. Calculation of the
phase relaxation rate for the case of a shallow donor results in the
expression, which is similar to Eq.~(\ref{eq11}),
\begin{eqnarray}
T_{2,0}^{-1} &=&\frac{\xi _{hf}\tau _{ph}^{-1}}{3}\sqrt{1+\frac{T^{2}}{%
T_{L}^{2}}},  \label{eq15} \\
\xi _{hf} &=&\frac{I(I+1)n_{I}a_{hf}^{2}}{6\pi ^{2}\hbar \rho
V_{QD}^{hf}c_{L}^{3}a_{0}^{2}}.
\end{eqnarray}
Similarly, one can find the approximate rate for an electron localized in a
QD through an analogy with Eq.~(\ref{eq12}),
\begin{equation}
T_{2,0}^{-1}=\xi _{hf}\tau _{ph}^{-1}\sqrt{c_{hf}^{2}+d_{hf}^{2}\frac{T^{2}}{%
T_{L}^{2}}},  \label{eq17}
\end{equation}%
where $c_{hf}=3.7-68\epsilon ^{2}$, $d_{hf}=2.7-9.8\epsilon ^{2}$, and $%
\epsilon \lesssim 0.1$. Numerical estimations provided for a donor in Si and
GaAs in terms of Eq.~(\ref{eq15}) indicate inefficiency of this mechanism
with a very long relaxation time (about 10$^{14}$~s and 10$^{8}$~s,
respectively). Hence, this mechanism can be neglected in most cases.

Along with the Hamiltonian given in Eq.~(\ref{r7}), the spin-two-phonon
interaction $H_{SL}^{(2)}=\sum D_{ijklmn}\mu _{B}B_{i}s_{j}u_{kl}u_{mn}$ can
also contribute to spin decoherence ($D_{ijklmn}$ is the spin-two-phonon
coupling constants). Now the fluctuating effective field takes the form $%
\Omega _{\alpha }=\sum_{p,p^{\prime }}W_{\alpha }^{p,p^{\prime
}}Q_{p}Q_{p^{\prime }}$, where $W_{\alpha }^{p,p^{\prime }}$ are the matrix
elements of $H_{SL}^{(2)}$. Hence the Fourier image of the correlation
function $\gamma _{\mu \nu }\left( \omega \right) $ [Eq.~(\ref{f12})] is
expressed in terms of phonon correlation functions $\left\langle
(Q_{p_{1}}Q_{p_{2}})\left( \tau \right) Q_{p_{3}}Q_{p_{4}}\right\rangle
_{\omega }$. Its calculation performed in a harmonic approximation leads to
a simple approximation for the spin phase relaxation rate
\begin{equation}
T_{2,0}^{-1}=\frac{\mu _{B}^{2}B^{2}D^{2}}{21\rho ^{2}c_{T}^{3}}\left( \frac{%
k_{B}T}{\hbar c_{T}}\right) ^{7}.  \label{eq18}
\end{equation}%
The parameter $D$ can be estimated as $D=3(g-2)C^{2}/E_{g}^{2}$ ($g$, $C$,
and $E_{g}$ are the electron $g$ factor, deformation potential and energy
gap). Numerical evaluation of Eq.~(\ref{eq18}) at low temperatures ($T=4$ K)
predicts a long relaxation time. In the case of GaAs and $B=1$~T, one can
find $T_{2,0}\approx 3\times 10^{5}$~s that is too long to be of any
experimental or practical interest.

\section{Precession fluctuations}

This section analyzes the elastic process listed as "Local field steps" in
Table I. The specific spin-phonon interactions under consideration are the
indirect mechanisms resulting from "g-tensor fluctuation" and "hyperfine
field fluctuation" (see Table II) via the stochastic transitions between
different electronic states.

\subsection{Qualitative consideration}

Consider an electron spin $\mathbf{s}$ under the influence of a magnetic
field directed along the $z$ axis whose strength fluctuates in a step-wise
fashion in time (see Fig.~4). In this case, the projection of electron spin
on the $z$ axis $s_{z}$ is conserved and no longitudinal relaxation occurs.
Nevertheless, the phase of electron spin will change randomly with the
Zeeman frequency fluctuation $\delta \Omega $ resulting in a decoherence
rate of $T_{2}^{-1}\sim \delta \Omega ^{2}\tau _{c}$. Here, $\tau _{c}$ is
the correlation time of the fluctuation.~\cite{Sem03} As discussed earlier,
the origin of $\delta \Omega $ can be the phonon-mediated stochastic
transitions between different orbital states. Since the spin splitting
characteristics vary with the orbital states, electrons experience
uncontrollable changes in the precession frequency (thus, the phase) while
undergoing spin-independent phonon scattering between them. In the case of a
QD, this process may play an important role owing to the shallow energy
levels and the dependence of $g$-tensor or hyperfine constant on the
electronic orbital states.~\cite{SKprl04}

\subsection{Theoretical model}

We begin the quantitative analysis by defining the Hamiltonian $H$ over the
basis functions consisting of a few lowest electronic states $\left\vert
n\right\rangle $, which are involved due to the phonon-assisted transitions.
We also consider that the single-electron problem in the QD gives a doubly
degenerate energy spectrum $E_{n}$ with the eigenstates $\left\vert
n\right\rangle $ in the absence of the Zeeman energy and the hyperfine
interaction. When a magnetic field is applied, it is conveniently assumed
that the differences in the spin splitting of the electronic states $%
\left\vert n\right\rangle $ and $\left\vert n^{\prime }\right\rangle $ are
small with respect to the energy intervals $\left\vert E_{n}-E_{n^{\prime
}}\right\vert $. The specific nature and type of the QD is unimportant for
the analysis. With these conditions, the total Hamiltonian can be expressed
as
\begin{equation}
H=H_{s}+H_{e}+H_{ph}+H_{e-ph}.  \label{q1}
\end{equation}

The first term $H_{s}$ is the spin (or pseudospin) energy Hamiltonian which
can be reduced to the form $H_{s}=\widehat{\mathbf{\Omega }}\mathbf{s}$ in
the most general case with $\widehat{\Omega }_{z}=\left\langle \uparrow
\left\vert H_{s}\right\vert \uparrow \right\rangle -\left\langle \downarrow
\left\vert H_{s}\right\vert \downarrow \right\rangle $ as discussed in Eq.~(%
\ref{f1}). Its projection on the lowest electronic states $\left\vert
n\right\rangle $ reads (see, for comparison, Refs.~\onlinecite{Hasegawa} and %
\onlinecite{Roth})%
\begin{equation}
\widehat{\mathbf{\Omega }}=\sum\limits_{n,n^{\prime }}\left\vert
n\right\rangle \mathbf{\Omega }^{n,n^{\prime }}\left\langle n^{\prime
}\right\vert ,  \label{q2}
\end{equation}%
where $\mathbf{\Omega }^{n,n^{\prime }}$ are the matrix elements of the
effective field (in units of energy) taken between the $\left\vert
n\right\rangle $ and $\left\vert n^{\prime }\right\rangle $ states. The
spin-independent (i.e., orbital) electron energies describe the Hamiltonian $%
H_{e}$
\begin{equation}
H_{e}=\sum\limits_{n}E_{n}\left\vert n\right\rangle \left\langle
n\right\vert .  \label{q3}
\end{equation}%
The Hamiltonians of the lattice and electron-phonon interactions have the
usual form%
\begin{eqnarray}
H_{ph} &=&\mathop{\displaystyle\sum}\limits_{p}\omega _{p}(a_{p}^{+}a_{p}+{%
\frac{1}{2}});  \label{q4} \\
H_{e-ph} &=&\mathop{\displaystyle\sum}\limits_{p,n,n^{\prime
}}B_{n,n^{\prime }}^{p}\left\vert n\right\rangle \left\langle n^{\prime
}\right\vert (a_{p}^{+}+a_{-p}).  \label{q5}
\end{eqnarray}%
%
%
%
%
Here, $B_{k,k^{\prime }}^{p}$ is the matrix element of the electron-phonon
interaction that depends on the material parameters and the geometry of the
QD. The last three terms of Eq.~(\ref{q1}) constitute the Hamiltonian of the
dissipative sub-system $H_{L}$ ($=H_{e}+H_{ph}+H_{e-ph}$) following the
notation of Eq.~({\ref{f1}). }

Thus, the problem of spin relaxation is reduced to calculating the
correlation functions of the effective field operator with the Hamiltonian $%
H_{L}$ of the dissipative subsystem. Clearly, the derivation depends
strongly on the specific form of $H_{L}$, the energy spectrum, and the
quantity of electron states considered. Keeping this context in mind, we
consider the simpler problem of electron fluctuations between only two
discrete states $\left\vert n\right\rangle =\left\vert g\right\rangle $ or $%
\left\vert e\right\rangle $ that are the ground and the first excited
electronic energy states (with an energy separation of $\delta _{0}$). The
corresponding Zeeman frequencies are denoted $\Omega _{z}^{(g)}$ [$=\Omega
_{z}^{g,g}$ in Eq.~({\ref{q2}})] and $\Omega _{z}^{(e)}$ ($=\Omega _{z}^{e,e}
$), respectively. Such a simplification allows us to easily perform all the
necessary calculations in an analytical form. In addition, most of the
important physics of the mechanism under consideration can be obtained in
the framework of this two-level model.

Hereinafter, it is convenient to introduce Pauli matrices $\sigma _{1}$, $%
\sigma _{2}$, $\sigma _{3}$ on the basis $\left\vert e\right\rangle $, $%
\left\vert g\right\rangle $, where according to the definition, $\sigma _{1}$%
, $\sigma _{2}$, $\sigma _{3}$ are invariant with respect to the coordinate
system rotation in contrast to actual spin matrices $\mathbf{s}$. Then, the
Hamiltonian of the dissipative subsystem takes the form
\begin{equation}
H_{L}=H_{ph}+\frac{1}{2}\delta _{0}\sigma _{3}+\Sigma _{p}B_{p}\sigma
_{1}(a_{p}^{+}+a_{-p}).  \label{q7a}
\end{equation}%
The electron spin Hamiltonian $H_{s}$ can now be split into the steady part $%
H_{S}$ and fluctuating component $H_{SL}$ with
\begin{equation}
H_{S}={\frac{1}{2}}(\mathbf{\Omega }^{e}+\mathbf{\Omega }^{g})\mathbf{s}%
\widehat{\mathbf{1}}  \label{q7b}
\end{equation}%
and
\begin{equation}
H_{SL}={\frac{1}{2}}(\mathbf{\Omega }^{(e)}-\mathbf{\Omega }^{(g)})\mathbf{s}%
\sigma _{3},  \label{q7c}
\end{equation}%
where $\widehat{\mathbf{1}}$ represents the unity matrix in the basis $%
\left\vert e\right\rangle $, $\left\vert g\right\rangle $. According to Eq.~(%
\ref{f5}), we must find $V=H_{SL}-\left\langle H_{SL}\right\rangle =\mathbf{%
\Omega }\mathbf{s}$ from Eq. (\ref{q7c}) that determines $\mathbf{\Omega }={%
\frac{1}{2}}(\mathbf{\Omega }^{(e)}-\mathbf{\Omega }^{(g)})(\sigma
_{3}-\left\langle \sigma _{3}\right\rangle )$ with $\left\langle \sigma
_{3}\right\rangle =-\tanh \left( \delta _{0}/2k_{B}T\right) $. The
correlation functions in Eq.~(\ref{f12}) now takes the form
\begin{eqnarray}
\gamma _{\mu \nu }\left( \omega \right)  &=&{\frac{1}{4}}(\Omega _{\mu
}^{(e)}-\Omega _{\mu }^{(g)})(\Omega _{\nu }^{(e)}-\Omega _{\nu
}^{(g)})J_{\omega }(T);  \label{q8} \\
J_{\omega }(T) &=&\left\langle \left( \sigma _{3}\left( \tau \right)
-\left\langle \sigma _{3}\right\rangle \right) \left( \sigma
_{3}-\left\langle \sigma _{3}\right\rangle \right) \right\rangle _{\omega }.
\label{q9}
\end{eqnarray}

The function $J_{\omega }(T)$ can be obtained by using the double-time
Green's function $G(t,t^{\prime })=\left\langle \left\langle \sigma
_{3}\left( t\right) ;\sigma _{3}\left( t^{\prime }\right) \right\rangle
\right\rangle $ with $H_{L}$. The final expression takes the following form,%
\cite{SKprl04}
\begin{eqnarray}
J_{\omega }(T) &=&\frac{1-\left\langle \sigma _{3}\right\rangle ^{2}}{\pi
n(\omega )}\frac{\tau _{c}}{\omega ^{2}\tau _{c}^{2}+1};  \label{q10} \\
\tau _{c}^{-1} &=&2\pi \mathop{\displaystyle\sum}\limits_{p}\left\vert
B_{p}\right\vert ^{2}\left\langle Q_{p}(\tau )Q_{-p}\right\rangle _{\delta
^{\prime }},  \label{q10a}
\end{eqnarray}%
where the difference between $\delta ^{\prime }=\delta _{0}-{\frac{1}{2}}%
(\Omega _{z}^{(e)}-\Omega _{z}^{(g)})$ and $\delta _{0}$ will be ignored in
the subsequent consideration. In the harmonic approximation, one can find
\begin{equation}
\left\langle Q_{p}(\tau )Q_{-p}\right\rangle _{\delta ^{\prime
}}=(2N_{p}+1)\delta \left( \omega _{p}-\delta _{0}\right) ,  \label{q11}
\end{equation}%
with the phonon population factor $N_{p}=\left\langle
a_{p}^{+}a_{p}\right\rangle $ for mode $p$. The parameter $\tau _{c}$ has
the simple physical meaning of the correlation time caused by
phonon-assisted transitions between the $\left\vert g\right\rangle $ and $%
\left\vert e\right\rangle $ states.

\subsection{Temperature dependence of phase relaxation}

Actually, Eqs.~(\ref{q8})-(\ref{q11}) describe the problem under
consideration in a very general form. As a result, an analysis on the
temperature dependence can be provided even before the details of the
phonon-mediated mechanism that induces the effective field fluctuation are
determined. Specifically, the correlation time $\tau _{c}$ given in Eq.~(\ref%
{q10a}) can be written as $\tau _{\delta }\tanh \left( \delta
_{0}/2k_{B}T\right) $ with the aid of Eq.~({\ref{q11}}), where $\tau
_{\delta }=\left[ 2\pi \Sigma _{p}\left\vert B_{p}\right\vert ^{2}\delta
\left( \omega _{p}-\delta _{0}\right) \right] ^{-1}$ is the lifetime of the
excited electron state with respect to the transition to the ground state
through phonon emission in the limit $T\rightarrow 0$. Thus, for the $\omega
=0$ component (e.g., $\gamma _{zz}^{0}$), the temperature dependence of spin
relaxation is reduced to
\begin{equation}
J_{0}(T)=\frac{\tau _{\delta }}{\pi }F\left( \frac{\delta _{0}}{2k_{B}T}%
\right) ;~~~F\left( x\right) =\left( 1-\tanh ^{2}x\right) \tanh x.
\label{q12}
\end{equation}

Figure~5 shows the numerical evaluation of $J_{0}(T)$ as a function of $T$
assuming $\delta_0 = 1$~meV. A maximum is observed at a temperature near $%
k_{B}T=\delta _{0}$ with distinctive slopes on each side. The left of the
peak corresponds to the reduced hopping from the $\left\vert g\right\rangle $
to $\left\vert e\right\rangle $ state that decreases the difference $\mathbf{%
\ \Omega }^{(g)}-\left\langle \mathbf{\Omega }^{(g)}\right\rangle $ (or the
amplitude of fluctuations). Hence, in the limit $T\ll \delta _{0}$ the
effective field fluctuations are frozen and our mechanism becomes
ineffective as $F\left( \delta _{0}/2k_{B}T\right) \longrightarrow \exp
(-\delta _{0}/k_{B}T) $. The slow negative slope on the high temperature
side [i.e., $F\left( \delta _{0}/2k_{B}T\right) \longrightarrow \delta
_{0}/2k_{B}T$] arises due to the well-known effect of dynamical fluctuation
averaging, which becomes more pronounced with an increase in temperature.

\subsection{Effect of \textit{g}-factor fluctuation}

The general theory discussed above is applied to a specific mechanism of
phase relaxation, which stems from the hopping between the excited and
ground states with different $g$-factors. In the most general case, the
reason for such a difference is the $g$-factor dependence on the energy
separation between the discrete electronic levels and the nearest
spin-orbital split electronic band. For technologically significant III-V
compounds, where the interaction with the valence band edge determines the
deviation of electron $g$-factor from the free electron Land\`{e} factor $%
g_{0}\approx 2.0023$,~\cite{Kiselev98} one can find the amplitude of the
fluctuation $\Delta g=(dg/dE_{g})\delta _{0}$ with $dg/dE_{g}=(g_{0}-g)(%
\Delta _{so}+2E_{g})/E_{g}(\Delta _{so}+E_{g})$, where $E_{g}$ is the band
gap and $\Delta _{so}$ the spin-orbital splitting of the valence band; we
also assume the inequality $\delta _{0}\ll $ $E_{g}$. In the case of a Si
QD, $E_{g}$ is a splitting of the $\Delta $ point in the Brillouin zone.~%
\cite{Roth} 
Then, the final equation for the phonon-assisted rate of phase relaxation
caused by the fluctuations of Zeeman splitting is given by
\begin{equation}
T_{2,Z}^{-1}=\frac{(g_{0}-g)^{2}}{4g^{2}}\left( \frac{\delta _{0}(\Delta
_{so}+2E_{g})}{E_{g}(\Delta _{so}+E_{g})}\right) ^{2}\omega _{0}^{2}\tau
_{\delta }F\left( \frac{\delta _{0}}{2k_{B}T}\right) ,  \label{q13}
\end{equation}%
where $\hbar \omega _{0}=g\mu _{B}B$ . One can see that our mechanism
reveals a quadratic dependence of $T_{2,Z}^{-1}$ on the applied magnetic
field $B$ in contrast to the $\sim B^{5}$ dependence found in the previous
calculations of longitudinal spin-lattice relaxation through the direct
(one-phonon) processes.~\cite{Roth,Hasegawa,Khaetskii01,Tahan02,GK03}

An estimation of the excited state lifetime $\tau _{\delta }$ can be
performed in terms of a deformation potential interaction and a model of
lateral carrier confinement. The matrix element of the corresponding
electron-phonon interaction between the $\left\vert g\right\rangle $ and $%
\left\vert e\right\rangle $ states is provided in Ref.~\onlinecite{GK03} as $%
B_{q}=iC\sqrt{\hbar q/2\rho v_{\parallel }V_{0}}J_{osc}$, where $%
J_{osc}=J_{osc}(\mathbf{q})$ is a form factor, $C$ is the deformation
potential, and $\rho $, $v_{\parallel }$ and $V_{0}$ are the density,
longitudinal sound velocity, and volume of crystal, respectively. A
straightforward calculation of inverse lifetime results in the expression%
\begin{equation}
\tau _{\delta }^{-1}=\frac{C^{2}q_{\delta }^{3}\alpha }{32\pi ^{2}\hbar \rho
v_{\parallel }^{2}}\displaystyle\int\limits_{0}^{1}(1-z^{2})e^{-\alpha
(1-z^{2})}dz,  \label{tau}
\end{equation}%
where $q_{\delta }=\delta _{0}/\hbar v_{\parallel }$, $\alpha =\delta
_{0}/2m_{e}v_{\parallel }^{2}$, and $m_{e}$ is the lateral effective mass.

To show the efficiency of the mechanism under consideration, we calculate
the relaxation rates in terms of Eqs.~(\ref{q13}) and (\ref{tau}) for a GaAs
QD with $\delta _{0}=1$~meV and the magnetic field $B=1$~T as a function of
temperature (Fig.~6, curve 1). A similar calculation is provided for a Si QD
with $g_{0}-g_{\parallel }=0.0131$ and $g_{0}-g_{\perp }=0.0141$~\cite%
{Wilson61} (Fig.~6, curve 2). A comparison between the result of Fig.~6 and
the $T_{1}$ calculation~\cite{GK03,Golovach04} shows that the phonon-induced
$g$-factor fluctuation via excited states can control the phase relaxation
in a Si QD at $T\gtrsim 2$ K (i.e., $T_{2,Z}<T_{1}$) in spite of a small $%
g_{0}-g$ (see Sec. VII for a detailed quantitative analysis); in the case of
a GaAs QD, this elastic process prevails over the spin-flip transitions at $%
T\gtrsim 1$ K and $B\leq 1$T.

\subsection{Effect of nuclear field fluctuation}

A similar fluctuation arises in the nuclear hyperfine field when the
electron undergoes uncontrollable transitions between the orbital $%
\left\vert e\right\rangle $ and $\left\vert g\right\rangle $ states. As
shown in Eq.~({\ref{eq13}}), the hyperfine interaction is influenced by the
electronic envelope functions evaluated at the location of the nuclear
spins. Hence, the difference in $\Psi _{e}(\mathbf{r})$ and $\Psi _{g}(%
\mathbf{r})$ leads to a dispersion of the hyperfine field whose mean value
is given as $\delta \Omega _{n}=a_{hf}\sqrt{\frac{2}{3}I(I+1)\varkappa
n_{I}/V_{QD}}$. Here, $I$ and $n_{I}$ are the nuclear spin and its
concentration in a QD of volume $V_{QD}$ as discussed before; the
dimensionless parameter $\varkappa =V_{QD}\int \left( \left\vert \Psi
_{e}\left( \mathbf{r}\right) \right\vert ^{2}-\left\vert \Psi _{g}\left(
\mathbf{r}\right) \right\vert ^{2}\right) ^{2}d^{3}\mathbf{r}$ is equal to $%
9/16\pi $ within the approximations of Ref.~\onlinecite{GK03}. If we set $%
\Omega _{z}^{(e)}-\Omega _{z}^{(g)}=\delta \Omega _{n}$ in Eq.\ (\ref{q8}),
we readily find the following estimation%
\begin{equation}
T_{2,hf}^{-1}=\frac{\varkappa }{6}I(I+1)\frac{a_{hf}^{2}n_{I}}{V_{QD}}\tau
_{\delta }F\left( \frac{\delta _{0}}{2k_{B}T}\right) .  \label{q14}
\end{equation}%
This equation shows the independence of spin relaxation ($T_{2,hf}$) on the
magnetic field when the process is induced by the hyperfine interaction. An
estimation of Eq.\ (\ref{q14}) (with an appropriate averaging over $\mathrm{%
^{69}Ga}$, $\mathrm{^{71}Ga}$ and $\mathrm{^{75}As}$)~\cite{Lampel77} for a $%
\mathrm{GaAs}$ QD with a typical size of $L_{xy}=50$ nm, $L_{z}=5$ nm and $%
\delta _{0}=1$ meV gives $T_{2,hf}=8\times 10^{-5}$~s, $2\times 10^{-7}$~s,
and $1.7\times 10^{-8}$~s for $T=1$ K, 2 K, and 4 K, respectively. A similar
estimation performed for a Si QD with the same dimensions and temperatures
results in $T_{2,hf}=29$~s, $9\times 10^{-2}$~s and $6.5\times 10^{-3}$~s,
respectively. Clearly, Eq.~(\ref{q14}) can provide a dominant contribution
for spin relaxation at low fields as it does not depend on the applied
magnetic field. In the case of the Si QD, a detailed quantitative analysis
is provided in Sec. VII.

\section{Intervalley transitions}

In this section, we consider one particular but important case that is
applicable to the semiconductors with multiple equivalent energy minima. The
basic principle is the same as that discussed in the previous section, i.e.,
via the phonon-mediated stochastic transitions. The difference is that the
present case takes into account the intervalley transitions between the
so-called "valley-split" states. Although normally prohibited, these
processes become possible in the presence of point defects. Of the two
potential mechanisms, the hyperfine field fluctuation is suppressed due to
the equivalence of the involved valleys. However, the "$g$-tensor
fluctuation" mechanism maintains its significance as the degeneracy of these
states is lifted (i.e.,"valley split" with different energies) in
asymmetrical structures.

\subsection{Qualitative consideration}

Following the brief discussion given above, this mechanism can occur in
semiconductors with multiple equivalent energy minima. As a specific
example, a QD grown along the [001] (i.e., $z$) direction with six
equivalent minima near the $X$ point is considered (i.e., Si-like); the
cases with different crystallographic symmetries can be analyzed with a
similar treatment. Subsequently, the valley-orbital structure of the QD
assumes the ground state formed by two equivalent valleys [001] and [00$%
\overline{1}$] with the wave functions $\psi _{001}$ and $\psi _{00\overline{%
1}}$. Any center-asymmetrical potential $V(r)$ (with respect to the $z$
direction) removes the degeneracy over the equivalent valleys with a
valley-orbital splitting $\Delta E_{v-v}$, thus forming the even and odd
states $\psi _{\pm }=(\psi _{001}\pm \psi _{00\overline{1}})/\sqrt{2}$. In
the case of a triangular potential $V(r)$, an estimation predicts the
splitting of $\Delta E_{v-v}=l_{v-v}dV(r)/dz$ with $l_{v-v}\approx 0.5~%
\mathrm{\mathrm{\text{\AA } }} $ in Si.~\cite{Grosso96} Clearly, one can see
that a moderately asymmetrical confinement potential can result in an energy
splitting sufficient large to induce the "$g$-tensor fluctuation" under an
appropriate lifetime $\tau _{\delta }$. However, the phonon induced
transitions between the valley-orbital states with a different parity are
normally forbidden or sufficiently suppressed.~\cite{Smelyanskiy05}

This restriction is lifted when one considers a structure composed of solid
solutions. For example, the Ge atoms in a Si$_{1-x}$Ge$_{x}$ QD with $x\ll 1$
(that offers a promising candidate for solid-state QC~\cite{Yablonovitch})
can be treated as the point defects in the Si lattice. The phonon scattering
with point defects due to the lattice anharmonicity results in the
appearance of high-frequency harmonics that can transfer electrons between
the equivalent valleys [001] and [00$\overline{1}$] or between the
valley-orbital states with a different parity. Figure 7 presents a diagram
of the process responsible for these transitions. As schematically
illustrated, the third-order anharmonicity splits a phonon resonant with $%
\Delta E_{v-v}$ into two high-frequency virtual phonons. One of them is
"frozen" due to the static deformation surrounding the defect, while the
other phonon induces the electron to undergo an intervalley scattering. The
transfer between the states with an energy separation leads to the
fluctuation in the electron precession frequency as described in the
previous section.

\subsection{Analysis of intervalley transition}

The estimation of spin decoherence rate due to the transitions between the
"valley-split" states can be performed in terms of Eq.~(\ref{q13}) when the
lifetime $\tau _{\delta }$ is calculated taking into account the
valley-orbital structure of a QD. However, direct application of Eq.~(\ref%
{q11}) to Eq.~(\ref{q10a}) would resulting the vanishingly small matrix
elements $B_{q}$ with $\hbar \omega _{q}=\Delta E_{v-v}$ for the case of
intervalley transitions. In order to take into consideration the process
depicted in Fig.~7, one must supplement the Hamiltonian [Eq.~(\ref{q4})]
with terms accounting for the phonon scattering at the defects:\cite%
{Carruthers61}
\begin{eqnarray}
H_{3} &=&\sum_{q,q^{\prime }}F_{q,q^{\prime },q^{\prime
}-q}Q_{q}Q_{q^{\prime }},  \label{v1} \\
F_{q,q^{\prime },q^{\prime }-q} &=&\frac{\hbar }{2\rho V_{0}}\frac{%
C_{q,q^{\prime }}}{\sqrt{\omega _{q}\omega _{q^{\prime }}}},  \label{v1a}
\end{eqnarray}%
where the matrix element of phonon scattering is%
\begin{equation}
C_{q,q^{\prime }}=i\frac{g_{C}}{2a^{3}}\sum_{\mathbf{a}}(\mathbf{V}%
_{q^{\prime }-q}\mathbf{a})(\mathbf{e}_{q}\mathbf{a})(\mathbf{e}_{q^{\prime
}}\mathbf{a})(\mathbf{q}\mathbf{a})(\mathbf{q}^{\prime }\mathbf{a})[(\mathbf{%
q}-\mathbf{q}^{\prime })\mathbf{a}].  \label{v2}
\end{equation}%
Here $g_{C}$ is a constant, $\mathbf{a}$ the unit vectors connecting
neighboring lattice atoms, and $\mathbf{V}_{q}$ a Fourier transformation of
the static displacement $\mathbf{y}(\mathbf{r})$ caused by point defects.
Additionally, a more general expression is applied for the correlation
function (see, for example, Ref.~\onlinecite{Zubarev60} and %
\onlinecite{Pathak65}) in place of Eq.~(\ref{q11}) that is obtained in a
harmonic approximation:
\begin{equation}
\left\langle Q_{p}(\tau )Q_{-p}\right\rangle _{\omega }=\frac{1}{\pi }\frac{%
\left( 2N_{p}+1\right) \Gamma _{p}(\omega )}{\left( \omega ^{2}-\omega
_{p}^{2}\right) ^{2}/\omega _{p}^{2}+\Gamma _{p}^{2}(\omega )}.  \label{eq3b}
\end{equation}%
The phonon damping rate $\Gamma _{p}(\omega )$ must be calculated in terms
of the interaction specified by Eqs.~(\ref{v1})-(\ref{v2}). Following the
works of Refs.~\onlinecite{Pathak65} and \onlinecite{Krivoglaz}, it is given
as
\begin{equation}
\Gamma _{p}(\omega )=8\pi \sum_{q^{\prime }}\left\vert F_{p,p^{\prime
},p^{\prime }-p}\right\vert ^{2}[\delta (\omega -\omega _{p^{\prime
}})-\delta (\omega +\omega _{p^{\prime }})].  \label{v5}
\end{equation}%
in terms of the Hamiltonian $H_{ph}+H_{3}$ [Eqs.~(\ref{q4}) and (\ref{v1})].

For simplicity, we ignore the lattice anisotropy and consider the
displacement $\mathbf{y}(\mathbf{r})$ in the model of isotropic medium:\cite%
{Carruthers61,Krivoglaz}
\begin{equation}
\mathbf{y}(\mathbf{r})=\sum_{j=1}^{N_{D}}A_{D}\frac{\mathbf{r}-\mathbf{R}_{j}%
}{\left\vert \mathbf{r}-\mathbf{R}_{j}\right\vert ^{3}},  \label{v3}
\end{equation}%
where $A_{D}$ is a mismatch in the lattice volume between the host and
substitutional impurity atoms. As it plays the role of deformation charge,
this parameter (i.e., $A_{D}$) can be found from the analysis of electron
paramagnetic resonance linewidth.~\cite{Kustov86} The Coloumbic deformation
field resulting from Eq.~(\ref{v3}) gives a well-known expression
\begin{equation}
\mathbf{V}_{q}=4\pi i\frac{A_{D}}{V_{0}}\sum_{j=1}^{N_{D}}e^{i\mathbf{q}%
\mathbf{R}_{j}}\frac{\mathbf{q}}{q^{2}}.  \label{v4}
\end{equation}%
Then, substitution of Eqs.~(\ref{v1a}), (\ref{v2}) and (\ref{v4}) into Eq.~(%
\ref{v5}) and follow-up integration over the phonon spectrum in the Debye
model (i.e., $\omega _{p}\equiv \omega _{q,\varkappa }=c_{\varkappa }q$)
results in
\begin{equation}
\Gamma _{\varkappa ,q}(\omega )=\frac{27\pi }{16}n_{D}\Sigma _{\varkappa }%
\frac{A_{D}^{2}g_{C}^{2}}{\rho ^{2}c_{L}^{7}}\omega ^{3}\omega _{q},
\label{v6}
\end{equation}%
where $n_{D}$ is the concentration of point defects. The coefficient $\Sigma
_{\varkappa }$ in this equation takes into account the contributions of the
phonon modes with different polarizations. In the approximation of an
isotropic medium, one can find $\Sigma _{L}=S_{LL}+(c_{L}/c_{T})^{2}S_{LT}$,
where the sums $S_{\varkappa \varkappa ^{\prime }}=\left\langle \left\{
\sum_{\widehat{a}}(\mathbf{e_{q}^{\varkappa }}\widehat{a})(\mathbf{%
e_{q}^{\varkappa ^{\prime }}}\widehat{a})(\widehat{q}\widehat{a})(\widehat{q}%
^{\prime }\widehat{a})[(\widehat{q}-\widehat{q}^{\prime })\widehat{a}%
]^{2}\right\} ^{2}\right\rangle _{(4\pi )^{2}}$ account for the polarization
mixing; $\widehat{a}=\mathbf{a}/a$; $\widehat{q}=\mathbf{q}/q$; $%
<...>_{(4\pi )^{2}}$ means averaging over the directions of $\mathbf{q}$ and
$\mathbf{q}^{\prime }$. A numerical estimation for the diamond lattice gives
$S_{LL}=0.789$ and $S_{LT}=0.239$; hence, $\Sigma _{L}=1.29$ for Si.
Finally, the Fourier transformation of the correlation function [Eq.~(\ref%
{eq3b})] is obtained as
\begin{equation}
\left\langle Q_{p}(\tau )Q_{-p}\right\rangle _{\omega }=\frac{27}{16\hbar }%
(2N_{p}+1)n_{D}\Sigma _{\varkappa }\frac{A_{D}^{2}g_{C}^{2}}{\rho
^{2}c_{L}^{7}}\frac{\omega ^{3}}{\omega _{q}}.  \label{v6a}
\end{equation}%
for the phonons connecting the equivalent valleys.

The next step is to evaluate the matrix element of the intervalley
electron-phonon interaction [see Eq.~(\ref{q10a})]. It assumes the form%
\begin{equation}
B_{p}=\sqrt{\frac{\hbar }{2\rho V_{0}\omega _{p}}}(\Xi _{d}+\frac{1}{3}\Xi
_{u})q\delta _{\varkappa ,L}\Phi (\mathbf{q}),  \label{v7}
\end{equation}%
where $\Xi _{d}$ and $\Xi _{u}$ are the deformation potential constants and
the form factor $\Phi (\mathbf{q})=\left\langle \psi _{\mp }\left\vert e^{i%
\mathbf{q}\mathbf{r}}\right\vert \psi _{\pm }\right\rangle $ selects the
wave vectors close to the $k$-space separation $\Delta k_{v}$ of the two
minima. Assuming a QD with the lateral and transversal dimensions of $L_{xy}$
and $L_{z}$, one can obtain
\begin{equation}
\Phi (\mathbf{q})\approx \exp [-\frac{L_{xy}^{2}}{4}(q_{x}^{2}+q_{y}^{2})-%
\frac{L_{z}^{2}}{4}(q_{z}+\Delta k_{v})^{2}].  \label{v9}
\end{equation}%
The appropriate $\Delta k_{v}$ takes the value of $\approx 0.3\pi /a_{L}$ ($%
a_{L}$ the lattice constant) considering the Umklapp process. Substitution
of Eqs.~(\ref{v6}) and (\ref{v7}) into Eq.~(\ref{eq3b}) results in the
contributions of two different types. The first is a direct process and, as
mentioned earlier, has only a minor significance to $\tau _{c}$ with a
characteristic factor $\sim \exp [-\frac{L_{z}^{2}}{2}\Delta k_{v}^{2}]\ll 1$%
. On the other hand, the second type accounts for the electron interaction
with virtual phonons of energy $\omega _{q}\gg \Delta E_{v-v}$. This
contribution is calculated with the help of Eqs.~(\ref{v6a})-(\ref{v9}):
\begin{eqnarray}
\tau _{c} &=&\tau _{\delta }\tanh \frac{\Delta E_{v-v}}{2k_{B}T},
\label{v10} \\
\tau _{\delta }^{-1} &=&C_{iv}n_{D}\gamma ^{2}\frac{A_{D}^{2}\Xi ^{2}\Delta
E_{v-v}^{3}}{\hbar ^{4}\rho V_{QD}c_{L}^{5}}  \label{v11}
\end{eqnarray}%
When deriving Eq.~(\ref{v10}), an estimation $g_{c}=24\gamma \rho \overline{c%
}^{2}$ from Ref.~\onlinecite{Carruthers61} is used that defines the constant
$C_{iv}\simeq 210$.

\subsection{Spin decoherence due to intervalley transitions}

With the intervalley transition mediated by the point defects as described
above, we can now calculate the subsequent spin relaxation due to the
associated fluctuation in the local magnetic field. Evidently, the hyperfine
interaction does not take part in the process since the equivalent
valley-orbital states possess identical electronic envelope functions.
Hence, only the $g$-tensor fluctuations due to the valley split energy is
considered. As in Sec.~V, applying Eq.~(\ref{v10}) to Eq.~(\ref{q13}) solves
the problem under the condition $\tau _{c}\ll T_{2}$. However, it should be
noted that the formal calculation of $T_{2}$ using these equations violates
the required inequality condition when $\Delta E_{v-v}$ is sufficiently
small. Subsequently, the Markovian kinetic equations [Eqs.~(\ref{f16})-(\ref%
{f18})] are unsuitable. Actually the spin decoherence time can be
determinate by $\tau _{c}$ itself in the regime, where the result for $%
T_{2,Z}$ [see Eq.~(\ref{q13})] is much shorter than $\tau _{c}$, as
schematically illustrated in Fig.~8. Here, the spin relaxation can be
considered in two stages. The first is the "pending" period of approximately
$\tau _{c}$, during which the electron has not undergone the intervalley
scattering. Obviously, no change occurs in the electron spin phase. Once the
transition is made, the electron is subject to the effective magnetic field
(due to the difference in the $g$ tensor) and starts to acquire the phase
change. Within a typical time on the order of $\Delta T_{2}\simeq \hbar
/\left\vert \Omega ^{(e)}-\Omega ^{(g)}\right\vert $ ($\ll \tau _{c}$), the
electron fully loses the information regarding the initial state. The
intermediate case $T_{2}\approx \tau _{c}$ remains beyond the quantitative
consideration. For simplicity, we apply an interpolation $T_{dc}\approx
T_{2}+\tau _{c}$. The calculated decoherence time $T_{dc}$ for a Si$_{1-x}$Ge%
$_{x}$ QD is plotted as a function of $\Delta E_{v-v}$ (Fig.~9) and
temperature (Fig.~10). Two cases of Ge composition ($x=0.1$ and $x=0.02$)
are considered assuming the QD size of $L_{xy}=50$~nm and $L_{z}=5$~nm.
Clearly, the proposed mechanism can significantly increase the spin
relaxation rate in the mixed crystals. Even with the smaller "defect"
concentration of 2~\%, $T_{dc}$ can be reduced to the 10~msec range.
Particularly, this rate is very strongly dependent on the magnitude of the
splitting $\Delta E_{v-v}$ (in reference to the thermal energy). Hence, the
control of the valley splitting will be essential for long coherence. For
estimation, we assume $\Xi =8.6$~eV, $\rho =2.33$~g/cm$^{3}$, $%
c_{L}=8.43\times 10^{5}$~cm/s, $\gamma =0.56$,~\cite{Klemens93} and $%
C_{iv}\approx 210$. The deformation charge $A_{D}$ was found to be $%
1.2\times 10^{-25}$cm$^{3}$ for a Ge atom\cite{commA} in Si from Ref.~%
\onlinecite{Kustov86}.

\section{Comparison between different mechanisms}

To gauge the significance of the elastic processes, the relevant transversal
spin relaxation rates ($T_2^{-1}$) are obtained numerically in a Si QD (with
$5\times 50\times 50$ nm$^{3}$) as a function of temperature, magnetic
field, and the orbital energy separation. In addition to the three elastic
processes described in this paper (i.e., anharmonic vibration, local field
steps due to the $g$-tensor fluctuation and hyperfine field fluctuation),
the contribution by the direct spin flip (that is the dominant inelastic
process in Si) is taken into account for comparison.~\cite{GK03} Since Si is
not a mixed crystal and the defect density kept low for the QC application,
the process via the intervalley transition is not considered. Figure~11
shows the calculated $T_2^{-1}$ vs.\ $B$ when the energy separation $%
\delta_0 $ is fixed at 2 meV. For the range of temperatures considered in
the study, it is apparent from the result that the direct spin-flip
dominates at a sufficiently high magnetic field ($\gtrsim 1$~T) due to its
strong dependence (e.g., $B^4 $-$B^5$). For low fields, the anharmonic
vibration (i.e., phonon decay) process tends to be significant particularly
when the temperature is also low. As the temperature increases, the
contributions due to the local field steps become prominent. While the $g$%
-tensor fluctuation mechanism plays a greater role at a stronger $B$, the
field independent nature of the nuclear field makes it stand out in the
other extreme. The corresponding map of the dominant process/mechanism in
the $B$-$T$ parameter space ($\delta_0=2$~mev) is given in Fig.~12, clearly
signifying the importance of elastic relaxation.

A similar comparison is provided in Fig.~13 as a function of the orbital
energy separation $\delta_0$ at three different temperatures. The magnetic
field is set to 1~T. Note that the contribution due to the anharnomic
vibration process is not shown as its relaxation rate falls below the range
of presentation. From the figure, it is clear that the relaxation induced by
the local field steps (via both the $g$-tensor and hyperfine field
fluctuation mechanisms) are very sensitive to $\delta_0$; this can be
readily understood as the process requires a round-trip transitions by
overcoming $\delta_0$. Subsequently, its contribution is important only for
a small $\delta_0$ (in reference to $k_BT$). Since the spin-flip rate is
practically independent of $\delta_0$ and $T$ in the considered parameter
range, the cross-over point between these rates monotonically moves to a
higher $\delta_0$ with increasing $T$. By 4~K, the inelastic process drops
out of the picture. Of the two elastic mechanisms, the $g$ tensor case seems
to be relatively more efficient as $\delta_0$ increases. This may be due to
the fact that the change in the $g$ tensor is directly proportional to the
orbital energy separation. Figure~14 summarizes the dominant
process/mechanism identified in the $\delta_0$-$T$ parameter space at $B =1$%
~T. Considering the conditions typical for QC using an electron spin in a Si
QD (e.g., $\delta_0 \lesssim 1$~meV, $T \lesssim 1$~K, $B \gtrsim 1$~T), it
is expected that $T_2$ over 10 sec can be achieved. If a SiGe QD is used in
place of Si, then the decoherence process can become more active depending
on the valley-split energy. However, $T_2$ on the order of 10~msec or longer
is still attainable. The spin relaxation in a GaAs QD is predicted to be
much faster.

Table~III summarizes the functional dependence of the dominant relaxation
processes considered in this study. The zero-th power (e.g., $T^0$, $B^0$, $%
\delta_0^0$) denotes the independent nature to the corresponding parameter.
The unique set of $T$-, $B$- and $\delta _{0}$-dependence will facilitate
the experimental identification of each process. As the thermal energy
provides the unwanted error/noise, the parameter space relevant to QC is
where $T$ is much smaller than the other characteristic energies.

\section{Conclusion}

A group of elastic spin relaxation processes that do not involve the energy
exchange between the Zeeman subsystem and the thermal reservoir, is studied
in semiconductor QDs. Contrary to the common perception, these processes can
play a dominant role in the electron spin decoherence under the certain
conditions (magnetic field, temperature, orbital energy separation, etc.) as
identified in the investigation. Particularly, the calculation results
illustrate the potential significance of an elastic decoherence mechanism
originating from the intervalley transitions in semiconductor quantum dots
with multiple equivalent energy minima (e.g., the $X$ valleys in SiGe). The
detailed understanding obtained in this work will help optimizing the QD
based QC systems for long quantum coherence; for example, the Si based QDs
can offer a spin relaxation time that is much longer than the GaAs
counterparts. However, an additional, comprehensive effort is needed for a
complete analysis of spin decoherence taking into account all the possible
contributions and their manifestations.

\begin{acknowledgments}
This work was supported in part by the Defense Advanced Research Projects
Agency and the SRC/MARCO Center on FENA.
\end{acknowledgments}

\clearpage
\begin{table}[tbp]
\caption{Main contributions to electron spin dephasing (with T$_{2}^{\ast }$%
) in a QD spin based QC system. The "random local fields" category denotes
the processes that lead to phase diffusion of the multi-qubit system without
causing decoherence in the individual qubits (i.e., spin). The decoherence
or transversal relaxation ($T_2$) has contributions from both elastic and
inelastic processes for a single spin. The latter involves spin flip
transitions leading to the energy (or longitudinal) relaxation of the spin
states (i.e., $T_1$). The \# sign denotes the spin relaxation processes
induced by the spin-phonon interactions. HFI symbolizes the hyperfine
interaction.\newline
}%
\begin{tabular}{|c|c|c|}
\hline
\multicolumn{3}{|c|}{\textbf{Dephasing (T}$_{2}^{\ast }$\textbf{)}} \\ \hline
\multicolumn{2}{|c|}{\textbf{Decoherence (T}$_{2}$\textbf{)}} & \textbf{%
Random Local} \\ \cline{1-2}
\textbf{Inelastic process (T}$_{1}$\textbf{)} & \textbf{Elastic process} &
\textbf{Fields} \\ \hline
Direct spin-flip$^{\#}$ & Anharmonic vibration$^{\#}$ & \textit{g-}factor
dispersion \\
Orbach$^{\#}$ & Local field steps$^{\#}$ & Dispersion of HFI \\
Two-phonon$^{\#}$ & Spectral diffusion & Impurity fields \\
Nuclear spins & Geometric phase$^{\#}$ & Inter-qubit interaction \\ \hline
\end{tabular}%
\end{table}

\clearpage
\begin{table}[tbp]
\caption{Various mechanisms of the spin-phonon interaction in a manner
consistent to both elastic and inelastic processes listed in Table~I (marked
by \#). SOI and HFI denote the spin-orbit and hyperfine interactions,
respectively.\newline
}%
\begin{tabular}{|c|c|c|c|}
\hline
& \multicolumn{3}{|c|}{\textbf{Spin-Phonon Interaction Mechanisms}} \\
\cline{2-4}
& \textbf{Spin-orbit interaction} & \textbf{Hyperfine interaction} & \textbf{%
Spin-spin interaction} \\ \hline
Direct & SOI modulation & HFI modulation & Waller \\
&  &  & Exchange modulation \\
Admixture & Kronig$-$Van Vleck & Spin state mixing & Spin state mixing \\
Indirect & g-tensor fluctuation & Hyperfine-field fluctuation & Spin-spin
fluctuation \\ \hline
\end{tabular}%
\end{table}

\clearpage
\begin{table}[tbp]
\caption{Functional dependence of the dominant spin-phonon processes
on the
relevant parameters considered in this study. The zero-th power (e.g., $T^0$%
, $B^0$, $\protect\delta_0^0$) denotes the independent nature to the
corresponding parameter. The temperature $T$, magnetic field $B$, and Zeeman
frequency $\protect\omega $ are expressed in units of energy. $T^{\ast }$
stands for a characteristic temperature, based on which the low and high
temperature regimes are defined. Eqs refer to the corresponding equations
given in the text.\newline
}%
\begin{tabular}{|c|c|ccc|cc|cc|}
\hline
Process & Eqs & \multicolumn{3}{|c|}{Temperature} & \multicolumn{2}{|c|}{
Magnetic field} & \multicolumn{2}{|c|}{Energy separation} \\ \cline{3-9}
&  & $T\ll $ & $T^{\ast }$ & $\ll T$ & $T\ll \omega $ & \multicolumn{1}{|c|}{%
$T\gg \omega $} & $T\ll \delta _{0}$ & \multicolumn{1}{|c|}{$T\gg \delta
_{0} $} \\ \hline
Direct spin-flip &  & $T^{0}$ & \multicolumn{1}{|c}{$\omega $} &
\multicolumn{1}{|c|}{$T^{1}$} & $B^{5}$ & \multicolumn{1}{|c|}{$B^{4}$} &
\multicolumn{2}{|c|}{} \\ \hline
Anharmonic vibration & \ref{eq12} & $T^{0}$ & \multicolumn{1}{|c}{$T_{eff}$}
& \multicolumn{1}{|c|}{$T^{1}$} & \multicolumn{2}{|c|}{$B^{2}$} &
\multicolumn{2}{|c|}{$\delta _{0}^{0}$} \\ \hline
$g$-tensor steps & \ref{q13} & $e^{-\delta_{0}/T}$ & \multicolumn{1}{|c}{$%
\delta_{0}$} & \multicolumn{1}{|c|}{$T^{-1}$} & \multicolumn{2}{|c|}{$B^{2}$}
& $e^{-\delta _{0}/T}$ & \multicolumn{1}{|c|}{$\delta _{0}$} \\ \hline
HF-field steps & \ref{q14} & $e^{-\delta_{0}/T}$ & \multicolumn{1}{|c}{$%
\delta_{0} $} & \multicolumn{1}{|c|}{$T^{-1}$} & \multicolumn{2}{|c|}{$B^{0}$%
} & $\delta_{0}^{2}e^{-\delta _{0}/T}$ & \multicolumn{1}{|c|}{$%
\delta_{0}^{-1}$} \\ \hline
\end{tabular}%
\end{table}

\clearpage
\begin{figure}[tbp]
\epsfig{figure=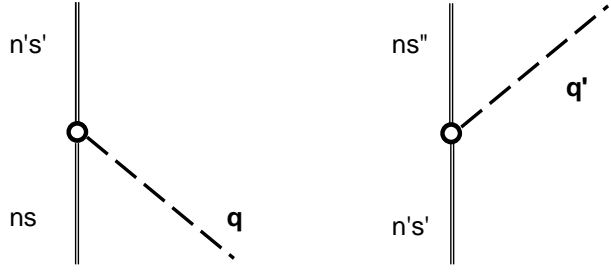,width=8cm,height=3.6cm,angle=0}
\caption{Schematic diagrams of one-phonon spin relaxation process.
The vertical lines correspond to the localized electrons with
orbital $n, n^{\prime} $ and spin $s=\uparrow $, $\downarrow $
states; the dashed lines depict a phonon with the $\mathbf{q}$ or
$\mathbf{q^{\prime}}$ mode; and the circle represents a direct
electron-phonon interaction. When the electronic transition due to
the interaction with a phonon does not change the orbital states
(i.e., $n=n^{\prime}$), the diagrams describe the direct process
of spin relaxation with the absorption (left panel) or the
emission (right panel) of the phonon resonant with the Zeeman
splitting. If $n$ and $n^{\prime}$ corresponds to the ground and
excited states respectively, the net effect of two depicted
processes in tandem represents the Orbach process of spin
relaxation. In this case (Orbach), the spin flip occurs only in
one of the stages (either absorption or emission) while the other
is spin independent.}
\end{figure}

\clearpage
\begin{figure}[tbp]
\epsfig{figure=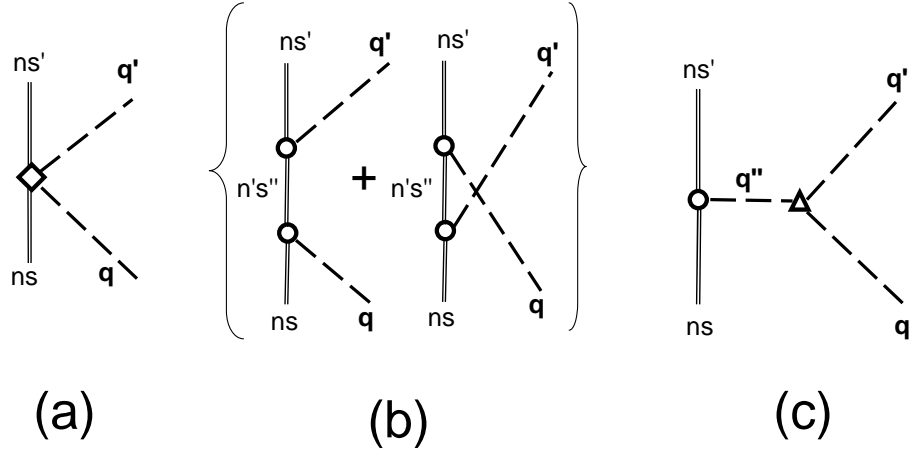,width=12cm,height=6cm,angle=0}
\caption{Schematic diagrams describing three types of spin flip $s
\rightarrow s^{\prime}$ Raman processes under the inelastic
two-phonon scattering. (a) The spin-two-phonon interaction in the
first order results in the direct Raman spin-relaxation process. (b)
The process of second order on the one-phonon interactions assumes
the electronic virtue transitions via intermediate states
$n^{\prime}s^{\prime \prime}$; the interference of the processes,
which corresponds to the braced diagrams, sometimes leads to the
so-called Van Vleck cancellation and the distinctive dependence
$\sim T^{9}$ for the relaxation rate. (c) The anharmonic spin
relaxation Raman process implies an inelastic phonon scattering due
to the third order anharmonicity with the creation of a virtual
phonon $q^{\prime\prime}$, which in turn induces a spin flip via the
spin-one-phonon interaction. Processes (a) and (c) are characterized
by typical temperature dependence $\sim T^{7}$ for the Raman spin
relaxation rate. As explained, the square represents the direct
spin-two-phonon interaction, while the triangle stands for the
inelastic phonon scattering due to the third order anharmonicity;
other notification are the same as in Fig.~1. }
\end{figure}

\clearpage
\begin{figure}[tbp]
\epsfig{figure=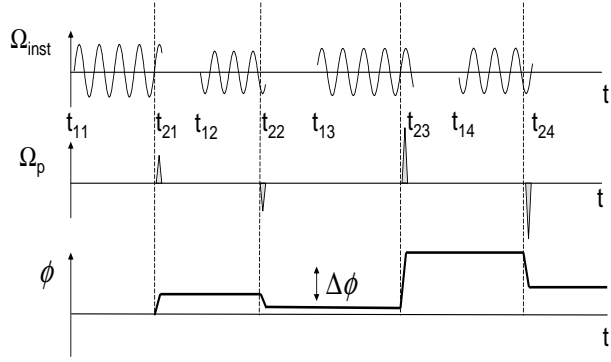,width=8cm,height=4.8cm,angle=0}
\caption{Schematic illustration of a single phonon contribution to
the random spin phase shift. $\Omega _{inst}$ is the instantaneous
effective field of spin precession associated with a phonon.
$t_{11}$, $t_{12}$, $t_{13}$, \ldots , are the random instants of
phonon generation, while $t_{21}$, $t_{22}$, $t_{23}$, \ldots , of
phonon annihilation. The vertical dotted lines depicts the
instants of the last full oscillations by the phonon before its
disappearance. The residual oscillations (on the right of the
dotted line) are not compensated and produce the net effect
($\Omega _{p}$), which influences the spin as a short pulses with
a random amplitude at random times. The phase shift due to the
influence of $\Omega _{p}$ is a Marcovian process at the time
interval $t\gg \protect\tau _{c}$, where the phonon correlation
time $\protect\tau _{c}$ is defined as a mean value of the
differences $t_{1i+1}-t_{1i}$ (thus, $\protect\tau _{c} =
\protect\tau _{p}$, the phonon relaxation time).  The phase
relaxation time is proportional to the mean value of the squared
phase shift caused by a single pulse $\left\langle \Delta
\protect\phi ^{2}\right\rangle $ and the rate of phonon relaxation
$\protect\tau _{p}^{-1}$.}
\end{figure}

\clearpage

\begin{figure}[tbp]
\epsfig{figure=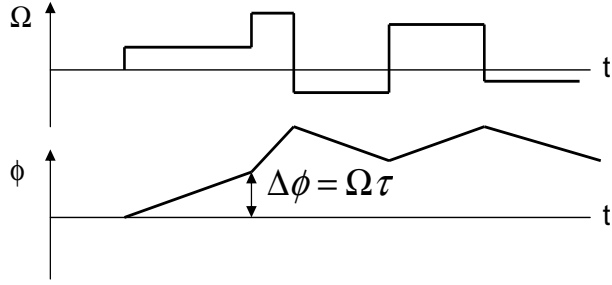,width=3.75cm,height=8cm,angle=270}
\caption{Schematic diagram illustrating the process of precession
fluctuations and their influence on spin phase relaxation. The
model assumes sudden changes of the effective field $\Omega$ at
random instances that lead
to the uncontrollable phase disturbances. At the scale of time $t\gg \protect%
\tau _{c}$ ($\protect\tau _{c}$ is the mean interval for consecutive quantum
leaps in $\Omega$), the Markovian process takes place that defines the
relaxation rate proportional to $T_{2}^{-1} \sim \left\langle \Delta \protect%
\phi ^{2}\right\rangle /\protect\tau _{c}$, where $\left\langle \Delta
\protect\phi ^{2}\right\rangle $ is the mean value of the squared phase
shift acquired between two successive sudden changes of $\Omega$. Taking
into account that $\Delta \protect\phi \approx \Omega \protect\tau _{c}$,
one can find the dependence $T_{2}^{-1} \sim \left\langle \Delta \protect%
\phi ^{2}\right\rangle \protect\tau _{c}$ typical for the process of
precession fluctuation.}
\end{figure}

\clearpage
\begin{figure}[tbp]
\epsfig{figure=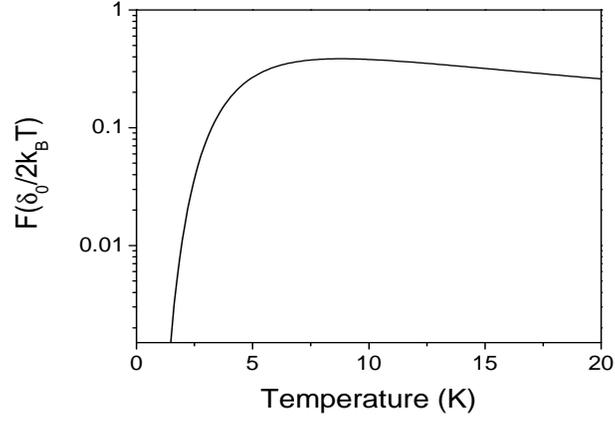,width=8cm,height=5.5cm,angle=0}
\caption{Decoherence factor $F(\protect\delta _{0}/2k_{B}T)$ [i.e., Eq.~(%
\protect\ref{q12})] for the process of precession fluctuation as a function
of temperature assuming $\protect\delta _{0}=1$ meV.}
\end{figure}

\clearpage

\begin{figure}[tbp]
\epsfig{figure=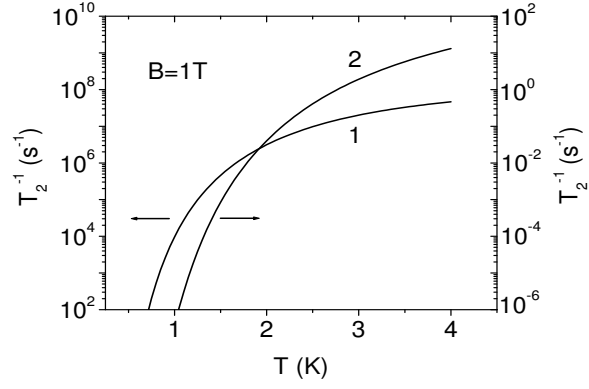,width=8.5cm,height=5.85cm,angle=0}
\caption{Spin decoherence rate via the $g$-tensor fluctuation
mechanism [i.e.,
Eq.~(\protect\ref{q13})] as a function of temperature for $\protect\delta%
_{0}=1$~meV and $B=1$~T. Curve 1 is for a GaAs QD, while curve 2 considers a
Si QD with $g_{0}-g_{\parallel }=0.0131$ and $g_{0}-g_{\perp }=0.0141$.}
\end{figure}

\clearpage
\begin{figure}[tbp]
\epsfig{figure=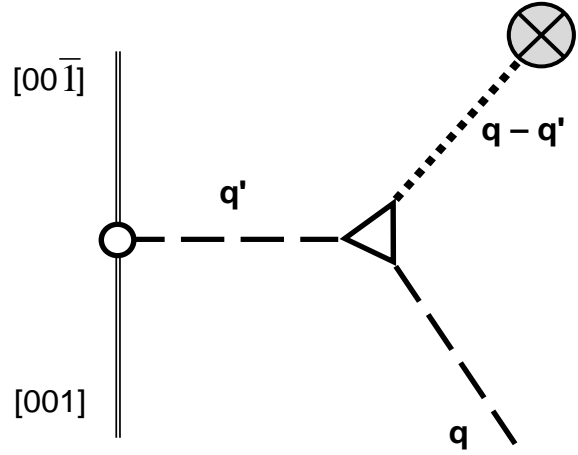,width=7.5cm,height=6cm,angle=0}
\caption{Schematic illustration of the phonon-mediated intervalley
relaxation process in a crystal with inhomogeneous deformations.
The $[001]$ and $[00\overline{1}]$ denote the electron states in
the two equivalent valleys; the circle represents the
electron-phonon matrix element, the triangle depicts third-order
anharmonicity, and the crossed circle is for the
point defect. The real phonon $\mathbf{q}$ splits into two virtual phonons $%
\mathbf{q}\prime $ and $\mathbf{q}-\mathbf{q}\prime $ due to the lattice
anharmonicity. The latter is accommodated by the local deformation of the
point defect that can be of short wavelength with $|\mathbf{q}\prime |\gg |%
\mathbf{q}|$. Consequently, the other virtual phonon can have a momentum $%
\mathbf{q}\prime $ large enough to induce the intervalley transition. }
\end{figure}

\clearpage

\begin{figure}[tbp]
\epsfig{figure=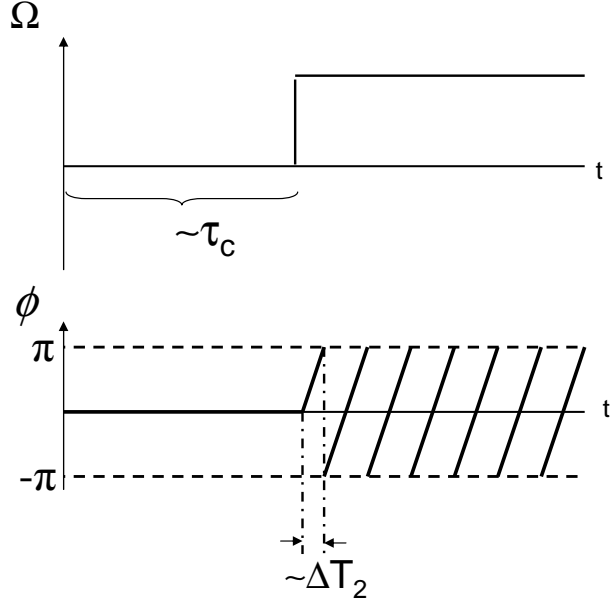,width=8cm,height=8cm,angle=0}
\caption{Diagram of phase relaxation in the case of inequality
$T_{2}\ll \protect\tau _{c}$, where $T_{2}$ is the transversal
relaxation time in the Markovian kinetic equations. In time of
approximately $\protect\tau _{c} $, the effective field is subject
to a sudden large change. This causes relatively fast spin phase
diffusion with the time scale of $\Delta T_{2}$. Hence, the duration
of the two-step process, $\approx \protect\tau _{c}$, can be
associated with the phase relaxation time.}
\end{figure}

\clearpage
\begin{figure}[tbp]
\epsfig{figure=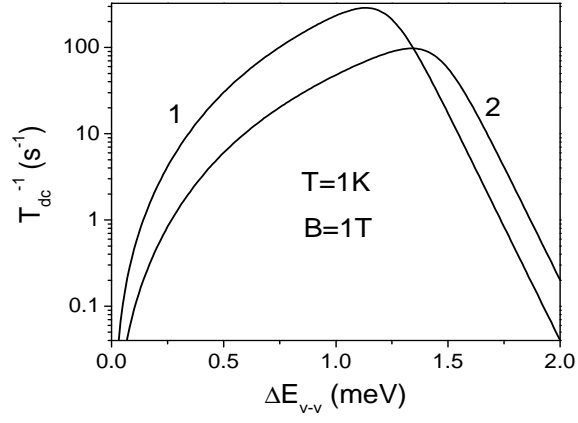,width=7.5cm,height=5.5cm,angle=0}
\caption{Spin decoherence rate via the intervalley transitions as
a function of valley-orbital splitting.  Curve 1 is for
Si$_{0.9}$Ge$_{0.1}$ (namely, the defect density of 10~\%), while
Si$_{0.98}$Ge$_{0.02}$ is considered for curve 2. }
\end{figure}

\clearpage
\begin{figure}[tbp]
\epsfig{figure=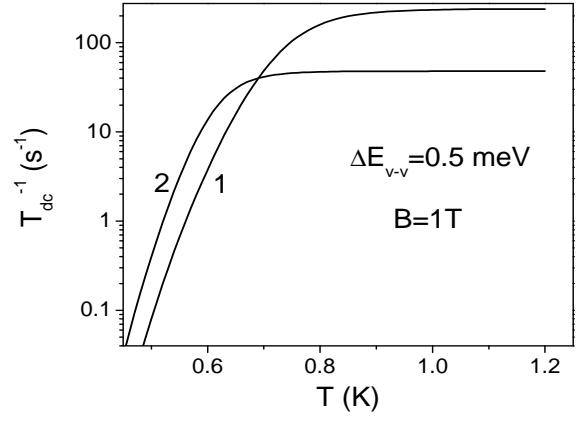,width=7.5cm,height=5.5cm,angle=0}
\caption{Spin decoherence rate via the intervalley transitions as
a function of temperature. Curve 1 is for Si$_{0.9}$Ge$_{0.1}$,
while Si$_{0.98}$Ge$_{0.02}$ is considered for curve 2.}
\end{figure}

\clearpage
\begin{figure}[tbp]
\epsfig{figure=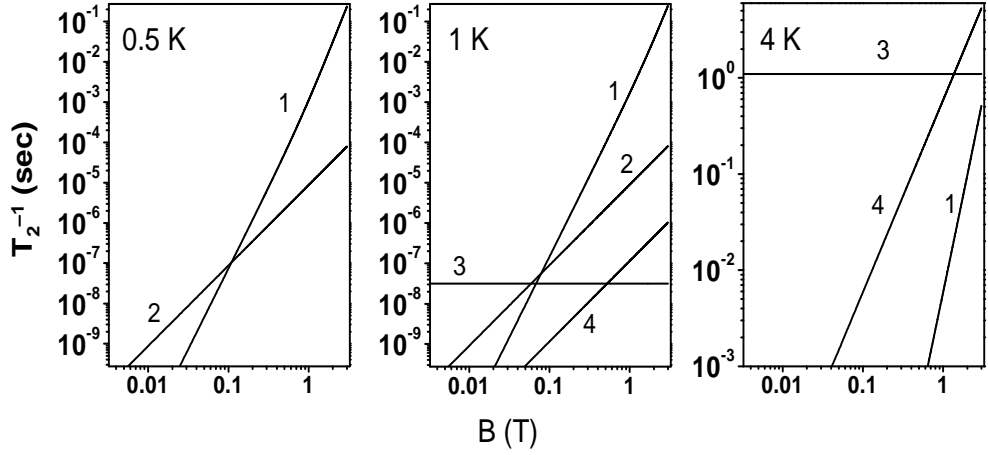,width=13cm,height=6cm,angle=0}
\caption{Spin decoherence rate vs.\ magnetic field for the
dominant spin-phonon relaxation processes in a typical Si QD (of
$5\times 50\times 50$ nm$^{3}$); line 1 $-$ inelastic direct
spin-flip process, line 2 $-$  elastic process of acoustic-phonon
anharmonic vibration, line 3 $-$ elastic process of local field
steps due to the hyperfine-field fluctuation, line 4 $-$ elastic
process of local field steps due to the $g$-tensor fluctuation.
The temperatures are as shown in the figure; the energy splitting
$\delta_0$  between the ground and the excited state is fixed at
2~meV. Since a Si QD is assumed without any point defects, the
process via the intervalley transition is not considered.}
\end{figure}

\clearpage
\begin{figure}[tbp]
\epsfig{figure=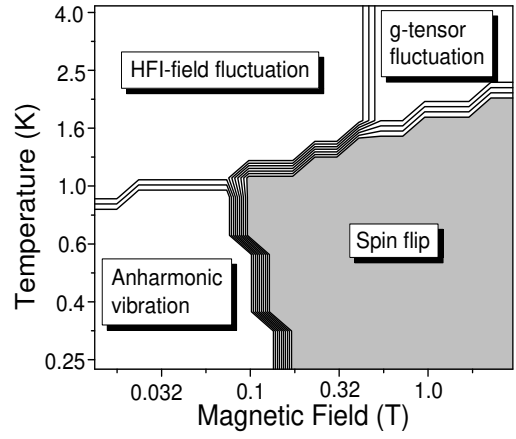,width=7.5cm,height=6.7cm,angle=0}
\caption{Phase diagram of dominant spin relaxation processes in
the Si QD in the $B$-$T$ parameter space.  The same conditions as
in Fig.~11 are assumed (for example, $\protect\delta _{0}=2$~meV).
The unshaded region is where the elastic spin-phonon processes
dominate.}
\end{figure}

\clearpage
\begin{figure}[tbp]
\epsfig{figure=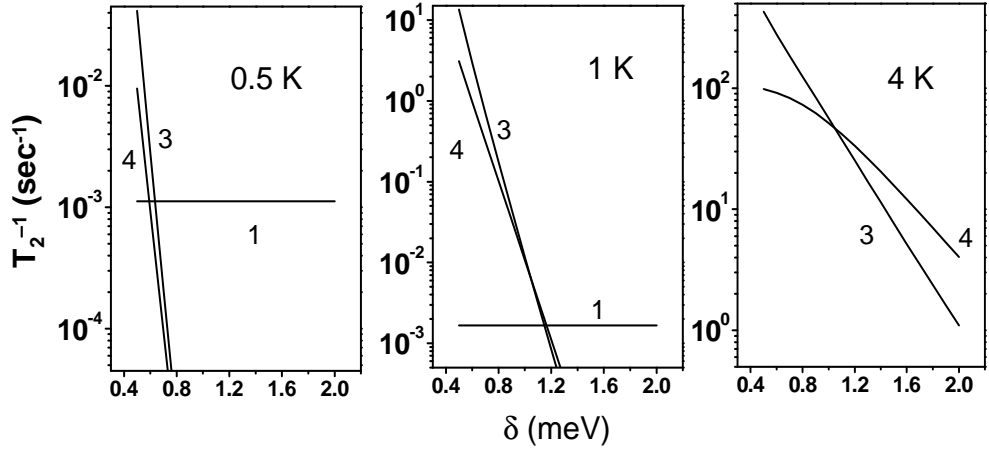,width=13cm,height=6cm,angle=0}
\caption{Spin decoherence rate vs. $\delta_0$  for the dominant
spin-phonon relaxation processes in a typical Si QD (of $5\times
50\times 50$ nm$^{3}$); line 1 $-$ inelastic direct spin-flip
process, line 3 $-$ elastic process of local field steps due to
the hyperfine-field fluctuation, line 4 $-$ elastic process of
local field steps due to the $g$-tensor fluctuation. The
temperatures are as shown in the figure; the magnetic field is
fixed at 1~T. Since a Si QD is assumed without any point defects,
the process via the intervalley transition is not considered.}
\end{figure}

\clearpage
\begin{figure}[tbp]
\epsfig{figure=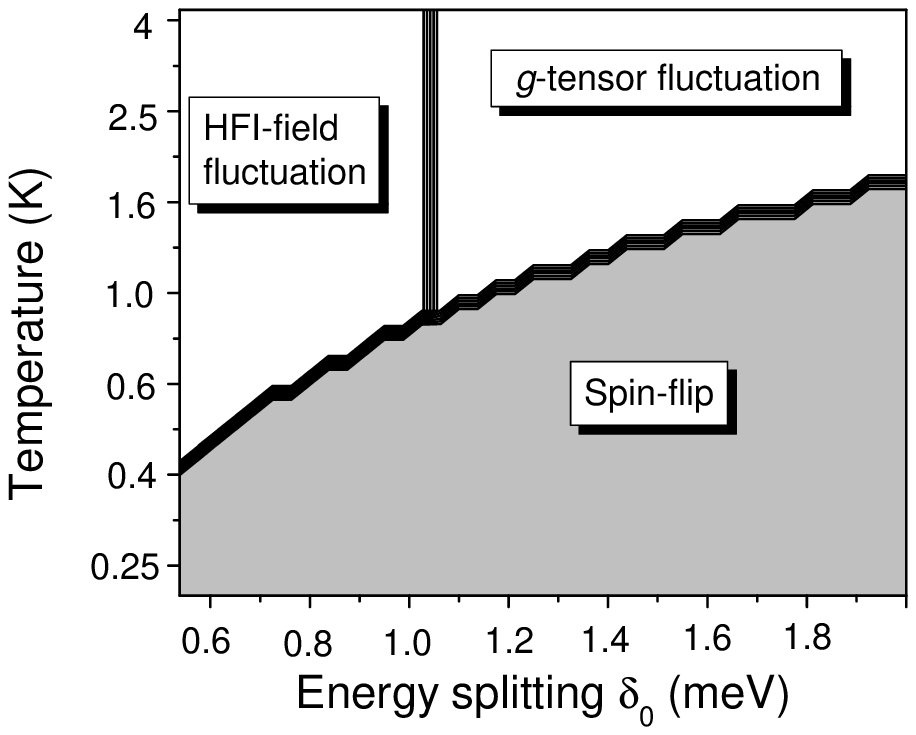,width=7.5cm,height=6.7cm,angle=0}
\caption{Phase diagram of dominant spin relaxation processes in
the Si QD in the $\delta_0$-$T$ parameter space. The same
conditions as in Fig.~13 are assumed (for example, $B=1 $~T). The
unshaded region is where the elastic spin-phonon processes
dominate.}
\end{figure}


\begin{references}
\bibitem{commA} Note that the approximation of the deformation charge $A$ as a
difference of volumes per Ge and Si atoms in the corresponding
crystals leads to an overestimation $A\simeq 3 \times
10^{-24}$~cm$^3$.
\end{references}
\end{document}